\documentstyle[aps,epsfig]{revtex}

\begin{document}
\draft
\def\be{\begin{equation}}
\def\ee{\end{equation}}
\def\bfi{\begin{figure}}
\def\efi{\end{figure}}
\def\bea{\begin{eqnarray}}
\def\eea{\end{eqnarray}}

\title{Scaling of the linear response function from zero field cooled and thermoremanent magnetization in phase ordering kinetics}

\author{Federico Corberi$^\dag$, Eugenio Lippiello$^\ddag$ 
and Marco Zannetti$^\S$}
\address {Istituto Nazionale di Fisica della Materia, Unit\`a
di Salerno \\ and Dipartimento di Fisica ``E.Caianiello'', 
Universit\`a di Salerno,
84081 Baronissi (Salerno), Italy}
\maketitle

\dag corberi@na.infn.it \ddag lippiello@sa.infn.it 

\S zannetti@na.infn.it

\begin{abstract}

In this paper we investigate the relation between the scaling properties of the linear
response function $R(t,s)$, of the thermoremanent magnetization (TRM) and of the 
zero field cooled magnetization (ZFC) in the context of phase ordering kinetics.
We explain why the retrival of the scaling properties of $R(t,s)$ from those of
TRM and ZFC is not trivial. Preasymptotic contributions generate
a long crossover in TRM, while ZFC is affected by a dangerous irrelevant variable.
Lack of understanding of both these points has generated some confusion in the
literature. The full picture relating the exponents of all the quantities involved is
explicitely illustrated in the framework of the large $N$ model. Following
this scheme, an assessment of the present status of numerical simulations for the
Ising model can be made. We reach the conclusion that on the basis of the data
available up to now, statements on the scaling properties of $R(t,s)$ can be  made
from ZFC but not from TRM. From ZFC data for the Ising model with $d=2,3,4$ we
confirm the previously found linear dependence on dimensionality of the exponent $a$ 
entering $R(t,s) \sim s^{-(1+a)}f(t/s)$.
We also find evidence that a recently 
derived form of the scaling function $f(x)$, using local scale 
invariance arguments 
[M.Henkel, M.Pleimling, C.Godr\`{e}che and J.M.Luck, Phys.Rev.Lett. {\bf 87}, 265701 (2001)], 
does not hold for the Ising model.

\end{abstract}

PACS: 05.70.Ln, 75.40.Gb, 05.40.-a

\section{Introduction}

The behavior of systems out of equilibrium is a subject of wide current 
interest~\cite{Cugliandolo2002}. Most of the attention is focused on glassy or disordered
systems. Nonetheless, many of the interesting features of slow relaxation, such as aging,
can be studied also in the simpler context of a phase ordering process.
This is the dynamical process which takes place, for instance, when a ferromagnet is
suddenly cooled from above to below the critical point. Then, ordered regions grow by coarsening.
The process is slow, i.e. the typical size of these regions grows with the power law
$L(t) \sim t^{1/z}$, where $z$ is the dynamic exponent. For dynamics with
non conserved order parameter (NCOP), as it will be considered in this paper, $z=2$
independent of dimensionality. In an infinite system equilibrium
is never reached. Phase ordering has been studied for a long time now~\cite{Bray94}. 
However, despite its relative simplicity when compared to the complexity
of glassy behavior, still there remains lack
of consensus and considerable confusion about the properties of the off equilibrium response function.
This paper is devoted to clarify the issue. This is a problem not of minor 
importance, given that phase ordering is regarded as a paradigmatic example of 
out of equilibrium behavior.

For definiteness, let us think of an Ising ferromagnet with
Hamiltonian
\be
{\cal H}[\sigma] =-J \sum_{<i,j>} \sigma_i \sigma_j
\label{01}
\ee
initially prepared at very high temperature and quenched at the time $t=0$ to a final temperature
$T < T_C$. In a process of this type the initial magnetization is zero and remains zero at all times
$\langle \sigma_i(t)\rangle =0$ for $t \geq 0$. Quantities of interest are~\cite{Bouchaud97} 
the autocorrelation function 
\be
C(t,s,t_0,t_{sc},t_{eq}) = \langle \sigma_i(t)\sigma_i(s)\rangle
\label{02}
\ee
where $t \geq s \geq 0$ are two times after the quench and the linear (auto)response function 
\be
R(t,s,t_0,t_{sc},t_{eq})= \left . {\partial  \langle \sigma_i(t)\rangle \over \partial h_i(s)} \right |_{h=0}
\label{03}
\ee
where $ h_i(s)$ is the external field conjugated to the order parameter.
Traditionally,
in phase ordering studies most of the attention has been devoted to the correlation
function~\cite{Bray94}, while the response function
has remained in the background.

In addition to
the two observation times $t$ and $s$, in Eq.s~(\ref{01}) and~(\ref{02}) we have explicitly
indicated also a dependence on the following characteristic times

\vspace{5mm}

\begin{itemize}

\item $t_0 \sim \Lambda^{-z}$. This is a microscopic time related, through the dynamic exponent $z$, to such a microscopic
length as the lattice spacing or the inverse momentum cutoff $\Lambda^{-1}$.

\item $t_{sc}$. The process of phase ordering is characterized by dynamical scaling in the asymptotic time
region (or late stage). The characteristic time  $t_{sc}$ separates the preasymptotic from the asymptotic regime,
i.e. it gives a measure of how much time is needed after the quench for scaling to set in.

\item $t_{eq}$.  After the formation of domains of ordered regions, equilibrium is rapidly reached in the
interior of domains. The characteristic time needed to establish this local equilibrium is the same as the
equilibration time in the pure ordered phases. It is given by  $t_{eq} \sim \xi^z$, where 
$\xi$ is the equilibrium correlation length in the pure phases at the final
temperature $T$ and $z$ is the dynamic exponent introduced above. 

\end{itemize}

\vspace{5mm}

\noindent The correlation and response function can always be written as the sum of two
contributions~\cite{Bouchaud97}
\be
    C(t,s,t_0,t_{sc},t_{eq})=C_{st}(t-s,t_0,t_{eq})+C_{ag}(t,s,t_0,t_{sc})
    \label{1.1}
    \ee
\be
    R(t,s,t_0,t_{sc},t_{eq})=R_{st}(t-s,t_0,t_{eq})+R_{ag}(t,s,t_0,t_{sc})
    \label{1.2}
    \ee
where the stationary contributions are what one has in equilibrium in the pure phases. Therefore, 
the usual fluctuation dissipation theorem is satisfied
\be
    R_{st}(t-s,t_0,t_{eq}) = \frac{1}{T}\frac{\partial C_{st}(t-s,t_0,t_{eq})}{\partial s}.
    \label{1.3}
    \ee
The rest, the aging contributions $C_{ag}(t,s,t_0,t_{sc})$ and $R_{ag}(t,s,t_0,t_{sc})$, 
are what is left over due to the existence of slow out of equilibrium degrees of freedom. 
The above split is useful for $s$ sufficiently large, i.e.
for
\be 
s \gg t_{eq}
\label{1.3.1}
\ee
in order to have well separated time scales for equilibrium and non equilibrium behavior 
and for 
\be
s \gg t_{sc}
\label{1.3.2}
\ee
in order for $C_{ag}(t,s,t_0,t_{sc})$ and $R_{ag}(t,s,t_0.t_{sc})$  to  
exhibit scaling behavior.

In connection with the aging contributions there are two basic questions

\vspace{3mm}

i) how do  $C_{ag}(t,s,t_0,t_{sc})$ and $R_{ag}(t,s,t_0,t_{sc})$ scale in the late stage

\vspace{3mm}

ii) what is the relation between $C_{ag}(t,s,t_0,t_{sc})$ and $R_{ag}(t,s,t_0,t_{sc})$, if any.

\vspace{3mm}

The second question belongs to the general area of the out of equilibrium
generalization of the fluctuation dissipation theorem~\cite{Crisanti2002}. 
This is a problem not as trivial as it is
believed to be for phase ordering systems~\cite{Corberi2001,Fusco2002}, 
with interesting implications
on the connection between statics and dynamics~\cite{Franz98}. 
In this paper we concentrate on the first question 
which is preliminary to the second one.

Assuming  that $s$ is large enough for (\ref{1.3.1}) and (\ref{1.3.2}) to be satisfied and dropping $t_{sc}$,  
the scaling form of $C_{ag}(t,s,t_0)$ is given by
\be
C_{ag}(t,s,t_0) \sim s^{-b}g(t/s,t_0/s).
\label{1.4}
\ee
It is well known~\cite{Bray94} that $b=0$. Furthermore, for $s \gg t_0$ one can set $y=0$ in $g(x,y)$ 
and it is also well known that for $x \gg 1$ one has   $g(x,0)=g(x) \sim x^{-\lambda/z}$, where $\lambda$ is the 
Fisher-Huse exponent.
Information about $R_{ag}(t,s,t_0)$, instead, is scanty. Writing the scaling relation analogous to Eq.~(\ref{1.4})
in the form
\be
    R_{ag}(t,s,t_0)=s^{-(1+a)}f(t/s,t_0/s)
    \label{1.5}
    \ee
both $a$ and $f(x,y)$ are much less known than $b$ and $g(x,y)$. 
Despite considerable efforts, no consensus has 
been reached as of yet on the value of $a$. The situation for the scaling function $f(x,y)$
is not much better. Recently, Henkel, Pleimling, Godr\`{e}che and Luck (HPGL)~\cite{Henkel2001},
using local scale invariance~\cite{LSI}, have derived an explicit form of the scaling function which is supposed to be 
of general validity. However, under close scrutiny this form appears neither to be obeyed in
those cases where an exact solution is available, nor to fit numerical data for
Ising systems, as it will be shown in section~\ref{zero}.

There is more than one reason for such an unsatisfactory state of affairs.
The first one is due to a qualitative analysis~\cite{Barrat98} of the relation between the
response function and the density of defects.
A naive use of this argument leads
to the conclusion that $a$ is independent of dimensionality, e.g. for scalar systems
$a=1/z$. In this form, due to its simplicity, this argument has become deeply
rooted in the literature~\cite{Franz98,Parisi,Berthier},
despite the accumulation of exact~\cite{Lippiello2000,Godreche2000,Corberi2002},
approximated~\cite{Corberi2001,Berthier}     and
numerical results~\cite{Corberi2001,CLZ2001,Corberi2003} incompatible with it. As we shall see,
$R_{ag}(t,s,t_0)$ is trivial in the sense that is proportional to the defect density 
only in the short time regime, but in no case this implies that $a$ is independent of
dimensionality.
Another reason is that in simulations $R_{ag}(t,s,t_0)$ is too noisy to work with and, in
order to deal with more manageable quantities, one must resort to the integrated response functions (IRF). 
The price for this is that reconstructing the scaling properties of  $R_{ag}(t,s,t_0)$
from those of an IRF is not as simple as it might look at first sight~\cite{Corberi2003}.
This will be the main theme of the paper.

We will show that, through the combined use of exact
results and numerical simulations, definite conclusions can be reached 
for the exponent $a$ by analysing in detail what actually goes on in the different methods
employed to evaluate it. For what concerns the scaling function $f(x,y)$,
instead, our understanding of the problem remains incomplete.

The paper is organized as follows. In section~\ref{what} we review existing information about
$R_{ag}(t,s,t_0)$, we make general considerations on the scaling function and we comment on the HPGL theory. 
In section~\ref{zero} we analyse the problem of retriving the properties of
$R_{ag}(t,s,t_0)$ from those of an IRF concentrating on the zero field cooled magnetization. 
Section~\ref{trm} is devoted 
to the same problem from the side of the thermoremanent magnetization. In section~\ref{larg} we use the
solution of the large $N$ model as an explicit illustration clarifying
what goes on when different IRF are employed to obtain information
on $R_{ag}(t,s,t_0)$. Concluding remarks are made in section~\ref{concl}.

\section{What is known about  $R_{ag}$ } \label{what}

This paper is devoted to the study of the exponent $a$ and the scaling function $f(x,y)$
entering Eq.~(\ref{1.5}). We first summarise what is known 
from exact and approximate analytical results providing direct access to  $R_{ag}(t,s,t_0)$.
We, then, make general considerations on $f(x,y)$ and some remarks on the HPGL form for it.

\vspace{5mm}

{\it Ising model d=1}

\vspace{5mm}
In the exact analytical computation of the response function~\cite{Lippiello2000,Godreche2000} 
in the  $d=1$ kinetic Ising model with Glauber dynamics, after taking $s \gg t_{sc}$ and 
neglecting $t_0/s$ one finds 
\be
R_{ag}(t,s) \sim s^{-1}(t/s-1)^{-1/2}
\label{2.6.1}
\ee
from which follows
\be
a=0
\label{2.6}
\ee
and
\be
f(x,0) \sim (x-1)^{-1/2}.
\label{2.6.0}
\ee
Furthermore, the correlation function is given by~\cite{Bray89,Prados97} 
\be
C(t,s) = {2 \over \pi} \arcsin \sqrt{{2 \over 1+t/s}}
\label{2.6.2}
\ee
which gives $C(t,s) \sim (t/s)^{-1/2}$ for $t/s \gg 1$. Hence, recalling $z=2$,
one has $\lambda=1$ and Eq.~(\ref{2.6.0}) can be rewritten as
\be
f(x,0) \sim \frac{x^{a+1/2-\lambda/z}}{ (x-1)^{a+1/2}}.
\label{2.7}
\ee
It should be mentioned that $a=0$ has been found numerically also in the case of the
kinetic Ising chain with Kawasaki dynamics \cite{Castellano2002}.

\vspace{5mm}

{\it Large N model}

\vspace{5mm}

Solving analytically the large $N$ model we
have found~\cite{Corberi2002} (see also section~\ref{larg}) $R_{ag}(t,s,t_0)$ of the form~(\ref{1.5})
with
\be
a=(d-2)/2
\label{2.10}
\ee
and
\be
f(x,y) \sim \frac{ x^{a+1-\lambda/z} -1}{ (x-1+y)^{a+1}}
\label{2.9}
\ee
where $d$ is arbitrary and $\lambda=d/2$. Notice that $a=0$ for $d=2$.

\vspace{5mm}

{\it Gaussian auxiliary field (GAF) approximation}

\vspace{5mm}

Berthier, Barrat and Kurchan~\cite{Berthier} have calculated analytically an IRF using
a GAF approximation based on the Ohta-Jasnow-Kawasaki method~\cite{Bray94}. From their
computation it is easy to extract $R_{ag}(t,s,t_0)$ which is in the form~(\ref{1.5}) with
\be
a=(d-1)/2
\label{2.13}
\ee
and
\be
f(x,y) \sim \frac {x^{a+1/2-\lambda/z}}{ (x-1+y)^{a+1/2}}
\label{2.12}
\ee
with $\lambda=d/2$.
Their calculation involves a diffusion constant of the form $D=(d-1)/d$
which prevents letting $d \rightarrow 1$, so they consider $d \geq 2$. 
We have worked out~\cite{Corberi2001} 
an alternative GAF approximation, without restriction on dimensionality,
which extends Eq.s~(\ref{2.13}) and~(\ref{2.12}) to $d \ge 1$. Then, we recover $a=0$ for
$d =1$ as in  the Ising case.

\subsection{General form of $f(x,y)$ and implications for $R_{ag}(t,s,t_0)$} \label{gen}

All the above results for $f(x,y)$ are of the form
\be
f(x,y) \sim \frac{ x^{-\beta} - \epsilon}{(x-1+y)^{\alpha}}
\label{2.15}
\ee
where $\epsilon =0$ if the correlation length in the low temperature pure phase is finite,
like in the $d=1$ Ising model and in the GAF approximation, or $\epsilon =1$ if the
low temperature phase is critical~\cite{note} like in the large $N$ model~\cite{Corberi2002} (see also section~\ref{larg}). 

We now make the phenomenological assumption that Eq.~(\ref{2.15}) is valid in general. Then the task 
becomes that of finding the exponents $a$, $\alpha$ and $\beta$. For this it is useful to look at the short
and long time behaviors.

\vspace{5mm} 

{\it Short time behavior}

\vspace{5mm}

\noindent Let us rewrite $R_{ag}(t,s,t_0)$ introducing the time difference $\tau=t-s$ in  Eq.~(\ref{2.15}) 
\be
R_{ag}(t,s,t_0) = s^{\alpha-(1+a)}\left [{(\tau/s +1)^{-\beta} -\epsilon \over (\tau +t_0)^{\alpha}} \right ].
\label{2.15.0}
\ee
Keeping $\tau$ fixed and letting $s$ to become large, to lowest order in $\tau/s$ we find
\be
R_{ag}(t,s,t_0) \sim s^{-\delta} \left [ \frac{\tau^{\epsilon}}{(\tau +t_0)^{\alpha}} \right ]
\label{10.2}
\ee
with
\be
\delta = (1+a)-(\alpha - \epsilon)
\label{10.3}
\ee
and where $\epsilon$ is the same as in Eq.~(\ref{2.15}). Therefore, from the short time
behavior one can extract $\delta$. An important observation is that
in the three explicit cases considered above $\delta$ coincides with
the exponent entering in the time dependence of the density of defects.
At the time $s$, this is given by 
\be
L(s)^{-n} \sim s^{-n/z}
\label{2.15.1}
\ee
where $L(s)$ and $z$ are the domain size and the dynamic exponent introduced above, 
$n=1$ for $N=1$, $n=2$ for $N>1$ and
$N$ is the number of components of the order parameter~\cite{Bray94}. One can, then, immediately verify that
\be
\delta = n/z.
\label{20.1}
\ee
In the $d=1$ Ising model and in the GAF approximation where $\epsilon =0$
and 
\be
\alpha = a+1/2
\label{50.2}
\ee
from Eq.~(\ref{10.3}) we get $\delta =1/2$, while in the large $N$ model with  $\epsilon =1$
and
\be
\alpha = a+1
\label{50.1}
\ee
we get $\delta =1$.

\vspace{5mm}

{\it Long time behavior}

\vspace{5mm}

\noindent In the large time regime $t/s \gg 1$, from Eq.s~(\ref{1.5}) and~(\ref{2.15}) follows 
\be
R_{ag}(t,s,t_0) \sim s^{-(1+a)} (t/s)^{-\lambda_R/z}
\label{30.1}
\ee
with
\be
\lambda_R/z =\alpha  +\beta.
\label{30.2}
\ee

\vspace{5mm}

\noindent Summarising,  the exponents $a$, $\alpha$ and $\beta$ can be obtained, in principle, by
making three different measurements on $R_{ag}(t,s,t_0)$: 

\begin{enumerate} 

\item $s$ dependence for fixed $t/s$ gives $a$ (from Eq.~(\ref{1.5}))

\item $s$ dependence for fixed $\tau$ gives $\delta$ (from Eq.~(\ref{10.2}))

\item $t$ dependence for fixed $s$ gives $\lambda_R/z$ (from Eq.~(\ref{30.1})).

\end{enumerate}

\noindent Before going into this, let us comment on the form of the scaling function derived by
HPGL in Ref.~\cite{Henkel2001}.

\subsection{Response function from local scale invariance}

Without making the separation~({\ref{1.2}) between stationary and aging components
and neglecting the dependence on $t_0$, HPGL assume that the full response function
$R(t,s)$ obeys the scaling form
\be
R(t,s) \sim s^{-(1+a)}f_{HPGL}(t/s).
\label{2.14.100}
\ee
Then, using local scale invariance arguments they make the prediction that in general,
for phase ordering, one has
\be
f_{HPGL}(x) \sim \frac {x^{a+1-\lambda/z}}{(x-1)^{a+1}}
\label{2.14}
\ee
where $\lambda$ is the Fisher-Huse exponent, provided there are no long range correlations in
the initial condition. In support of~(\ref{2.14}) they invoke the
exact solution of the spherical model~\cite{GL2000,Cugliandolo95,Zippold2000}, which is
equivalent to the large $N$ model, and numerical
simulations for the Ising model with d=2 and d=3. We make the following comments

\begin{itemize}

\item in the spherical or large $N$ model Eq.~(\ref{2.14}) ideed reproduces the full response function.
This coincides with the aging contribution~(\ref{2.9}) for 
$x \gg 1$ but not for $x \simeq 1$. This difference will turn out to be important (see section~\ref{larg}).

\item Eq.~(\ref{2.14}) is contained in the general form~(\ref{2.15}) with
$\epsilon =0$, $\beta= \lambda/z-(a+1)$ and $\alpha = a+1$. Inserting into Eq.~(\ref{10.3})
follows that in all cases one should have $\delta =0$. Furthermore, one should also have
$\alpha = a+1$ always, while from the explicit examples considered above this is true only
in the large $N$ case and not in the $d=1$ Ising model or in the GAF approximation.

\item HPGL theory is supposed to hold also for quenches at $T_C$. In that case the validity
of Eq.~(\ref{2.14}) has been questioned in the framework of the field theoretic 
$\epsilon$-expansion for the response function~\cite{Calabrese}.

\item About the support to Eq. (\ref{2.14}) 
from numerical simulations we will comment below. 

\end{itemize}

\vspace{5mm}

Now, in order to go beyond the explicitely solvable cases, 
the problem is to determine the exponents $a$, $\alpha$ and $\beta$ in 
the Ising model with $d >1$. As stated in the Introduction, measurement of $R_{ag}(t,s,t_0)$
is too noisy, so the program outlined above in subsection~\ref{gen} on the basis of Eq.s~(\ref{10.2}) and (\ref{30.1})
requires an unrealistically long computing time. In the next
section we discuss how to proceed with the help of IRF.

\section{Zero field cooled magnetization} \label{zero}

Indirect information on $R_{ag}(t,s,t_0)$ comes from numerical results on IRF. In general, an IRF is defined by
\be
\mu(t,t_2,t_1,t_0,t_{sc},t_{eq}) = \int_{t_1}^{t_2} ds R(t,s,t_0,t_{sc},t_{eq})
\label{3.1}
\ee
with $t \ge t_2 \ge t_1 \ge 0$ and using~(\ref{1.2}) one has
\be
\mu(t,t_2,t_1,t_0,t_{sc},t_{eq}) = \mu_{st}(t-t_2,t-t_1,t_0,t_{sc},t_{eq}) + \mu_{ag}(t,t_2,t_1,t_0,t_{sc}).
\label{3.2}
\ee
We will concentrate on the second contribution in the right hand side.
The reason for introducing an IRF is that the integration over $(t_1,t_2)$ lowers the
noise. However, if one has to resort to an IRF, there is the related problem of 
retriving the properties of $R(t,s)$ from it. This is not straightforward.
IRF usually employed are 

\begin{enumerate}

\item the thermoremanent magnetization (TRM)
\be
\rho(t,t_w,t_0,t_{sc},t_{eq}) = \mu(t,t_2=t_w,t_1=0,t_0,t_{sc},t_{eq})
\label{3.3}
\ee
obtained by looking at the response at the time $t$ to an external field acting in the interval $(0,t_w)$

\item the zero field cooled magnetization (ZFC)
\be
\chi(t,t_w,t_0,t_{sc},t_{eq}) = \mu(t,t_2=t,t_1=t_w,t_0,t_{sc},t_{eq})
\label{3.4}
\ee
obtained by looking at the response at the time $t$ when the field acts in the interval $(t_w,t)$.

\end {enumerate}

\noindent Both these quantities do have shortcomings. For TRM the problem is evident. The
integration starts at $t_1=0$, so preasymptotic contributions are
always included. The dependence on $t_{sc}$ cannot be neglected and this
turns out to make it particularly hard to extract the asymptotic behavior in
the cases of interest.

With ZFC there is not such a problem. Taking $t_w \gg t_{sc}$, and neglecting
$t_{sc}$ thereafter, one can be confident to be in the asymptotic region where scaling
holds. So, using~(\ref{1.5}) with $f(x,y)$ of the general form~(\ref{2.15})
and considering the case with $\epsilon = 0$, one has
\be
\chi_{ag}(t,t_w,t_0) = t_w^{-a} F(t/t_w,t_0/t_w)
\label{3.4.1}
\ee
with
\be
F(x,y) = x^{-a} \int_{1/x}^1  dz \frac{ z^{\beta +\alpha -(a+1)}}
{(1-z +y/x)^{\alpha}}.
\label{3.4.2}
\ee
The first observation is that if one seeks to determine $a$ from Eq.~(\ref{3.4.1}) 
by looking at the behavior of $\chi_{ag}$ as $t_w$ is varied
and $x=t/t_w$ is kept fixed, one must be aware that the $t_w$ dependence coming from $t_0/t_w$
may play a role. In other words $y=t_0/t_w$ may act as a dangerous irrelevant variable.
Namely, defining the exponent $a_{\chi}$ by
\be
\chi_{ag}(t,t_w) \sim t_w^{-a_{\chi}} \widehat{\chi}(x)
\label{3.4.3}
\ee
there may be a difference 
between $a$ in Eq.~(\ref{3.4.1}) and $a_{\chi}$ in Eq.~(\ref{3.4.3}). 
This depends on whether the integral in Eq.~(\ref{3.4.2})
diverges or not at the upper limit of integration as $y \rightarrow 0$. This, in
turn, depends on the value of $\alpha$. The second observation is that $\alpha$ can be extracted from the
large $x$ behavior of $F(x,y)$, as we shall see in the following. 
Instead, the task of extracting $\beta$ from Eq.~(\ref{3.4.2}) remains exceedingly complicated.

\subsection{The exponents $a$ and $a_{\chi}$}
 
The possibility that $a_{\chi}$ might not be identifiable with $a$, due to the presence
of $y=t_0/t_w$, can be checked explicitely in the large $N$ model~\cite{Corberi2002}
and in the GAF approximation~\cite{Corberi2001,Berthier}, where ZFC can be calculated with arbitrary $d$.
In both cases  there is a value $d_{\chi}$ of the dimensionality
such that $y$ is dangerous irrelevant above $d_{\chi}$.
This implies that $a_{\chi}$ coincides with $a$ for $d  < d_{\chi}$ and is given by Eq.s~(\ref{2.10})
and~(\ref{2.13}). Instead, $a_{\chi}$ is different from $a$ and is given by 
\be
a_{\chi}= \delta
\label{3.7}
\ee
for $d  > d_{\chi}$ with $\delta$ given by Eq.~(\ref{20.1}), which is independent of dimensionality. 
Logarithmic corrections appear at
$d  = d_{\chi}$, much in the same way as at the upper critical dimensionality in 
ordinary critical phenomena. The relation between  $a_{\chi}$ and $a$ in these two
models is given by 
\be
    a_{\chi} = \left \{ \begin{array}{ll}
        a  \qquad $for$ \qquad d < d_{\chi}  \\
        \delta  \qquad $with log corrections for$ \qquad d=d_{\chi} \\
	\delta   \qquad $for$ \qquad d > d_{\chi} 
        \end{array}
        \right .
        \label{4.1}
        \ee
with $\delta=1$ and $d_{\chi}=4$
in the large $N$ model and with $\delta=1/2$ and $d_{\chi}=2$ in the GAF approximation.
We emphasize that in these two solvable cases, Eq. (\ref{3.7}) holds only for $d > d_{\chi}$ where
$a_{\chi} \neq a$.

Next, from extensive numerical simulations~\cite{CLZ2001,Corberi2001,Corberi2003} 
of the Glauber-Ising model with $d=2,3,4$ we have measured $a_{\chi}$ obtaining 
data which are fairly well consistent (Fig.~\ref{fig.1}) with the phenomenological formula
\be
    a_{\chi}= \left \{ \begin{array}{ll}
        (d-1)/4 \qquad $for$ \qquad d < 3 \\
        1/2  \qquad $with log corrections for$ \qquad d=3 \\
        1/2 \qquad $for$ \qquad d > 3.
        \end{array}
        \right .
        \label{3.8}
        \ee
Since in the scalar case $\delta=1/2$, it is evident that the pattern~(\ref{4.1}) is followed
also in the Ising model with $d_{\chi}=3$.

We may, then, conclude that 
in all cases: exact, approximate and numerical $a_{\chi}$ is given by Eq.~(\ref{4.1})
and that, therefore, the exponent $a$ obeys the general formula
\be
a= \delta \frac{d-d_L}{d_{\chi}-d_L}
\label{4.4}
\ee
where $d_L$ is the dimensionality where $a=0$. According to this picture,
the distinction among the different systems comes through the values of
$\delta$, $d_{\chi}$,  $d_L$ (see Table~\ref{tabella}). 

\begin{center}
\begin{minipage}{4cm}
\begin{table}[htbp]
\begin{tabular}{|l|c|c|c|}
    \hline  & Ising & GAF & $N=\infty$ \\
    \hline $\delta$ & 1/2 & 1/2&  1 \\
    \hline  $d_L$   &  1  &  1 &  2 \\
    \hline  $d_\chi$ &  3  &  2 &  4 \\
    \hline
\end{tabular}
\caption{}
\label{tabella}
\end{table}
\end{minipage}
\end{center}

In this respect, notice that for $N=1$ both 
from simulations and from GAF one has $\delta = 1/2$ and $d_L=1$, while
there is a discrepancy between $d_{\chi} =3$ and $d_{\chi} =2$.
However, this is not worrysome. As explained in
Ref.~\cite{Corberi2001}, the dimensionality dependence of $a_{\chi}$ below $d_{\chi}$ takes
place because $d_{\chi}$ is the dimensionality below which minimization of magnetic
energy competes effectively with minimization of surface tension in driving interface
motion. Therefore, the balance of these two mechanisms is very sensitive to the treatment
of surface tension and it should not come as a surprise that from an uncontrolled
approximation, such as are those of the GAF type, a value of $d_{\chi}$ which differs
from the one observed in simulations is obtained. The shift from $d_{\chi}=3$ to
$d_{\chi}=2$ means that in the GAF approximation surface
tension is overestimated with respect to simulations.

\vspace{1.5cm}
\begin{figure}
\centerline{\psfig{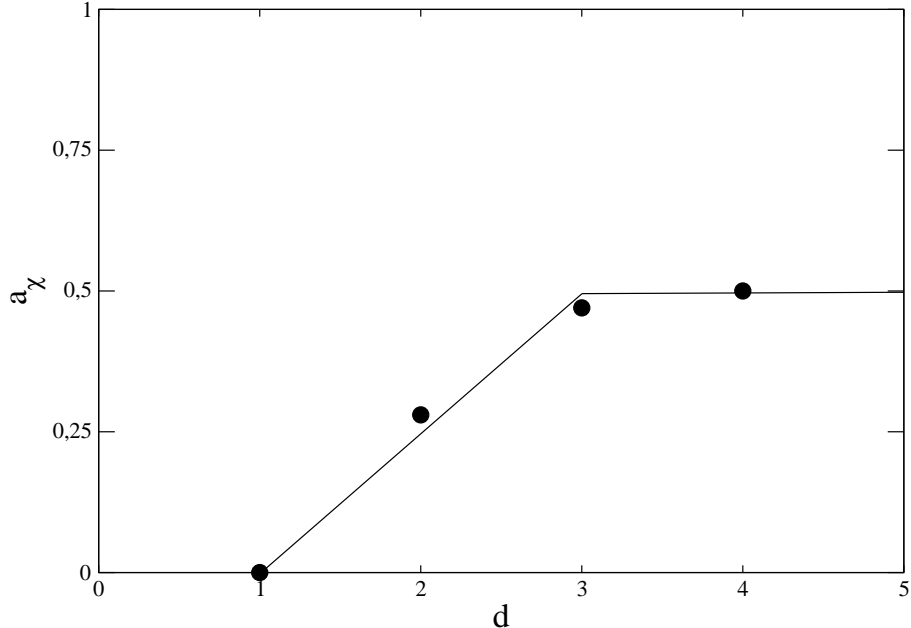}}
\caption{Exponent $a_{\chi}$ in the Ising model at 
various dimensionalities.
The continous line represents Eq.~(\ref{3.8}), while the dots are the values
from the exact solution of the model at $d=1$ and from simulations with
$d=2,3,4$.}
\label{fig.1}
\end{figure}

\subsection{The scaling function $\widehat{\chi}(x)$ and the exponent $\alpha$}

Although the results described above~\cite{CLZ2001,Corberi2001,Corberi2003}  
yield unequivocally $d_{\chi} =3$ for $N=1$,
in order to treat this point most carefully, we have investigated anew
the behavior of $\chi_{ag}(t,t_w)$ with very accurate simulations of the Ising model
with NCOP, $d=2,3,4$ and for different values of $t_w$ in order to get data also
on the scaling function $\widehat{\chi}(x)$, which has not been studied previously. 

First, let us illustrate the algorithm. There are several ways to isolate the aging 
contribution $\chi_{ag}(t,t_w)$.
The most obvious is to compute the total $\chi(t,t_w)$ by simulating a quenched
system and then to subtract from it the stationary part $\chi_{st}(t,t_w)$
obtained by simulation
of a system in equilibrium at the final temperature of the quench. 
A different algorithm was introduced by Derrida~\cite{Derrida97} regarding the
stationary contribution as due to thermal fluctuations inside the bulk of  
domains and the aging part as produced from the interfaces. The next 
step is to isolate the spins belonging to an interface. In order to
do this a  parallel simulation is performed of two systems with  different
initial conditions. The first is prepared in equilibrium at the initial temperature
and then is quenched to the final temperature $T$, 
while the second is in equilibrium at the final temperature $T$ from the beginning.
These two systems evolve with the same thermal history at the temperature $T$. 
At each time step spins that
are flipped in the first system but not in the second  are considered as interfacial
and their response is assigned to the aging part. 

These two methods are equivalent, but also numerically very inefficient. Let us
refer to them as global methods. The reason for inefficiency is that in order to
extract the response produced by the spins on the interfaces one has to simulate
the whole lattice. Since the interface density decreases as $t^{-1/z}$ a huge
amount of cpu time can be saved by an algorithm updating only the interfacial spins.
We stress that a fast algorithm is crucial in order to have reliable
results in a numerically hard problem such as this. Therefore, we have adopted 
a no bulk flip algorithm, where a list of interfacial spins is updated at each move
following the criterion that a spin belongs to an interface if at least one of
the nearest neighbourghs is not aligned. Only moves of the interfacial spins
are allowed. We then take the response of this system as $\chi_{ag}(t,t_w)$.  

In $d=1$ it can be shown~\cite{Lippiello2000} that
the no bulk flip algorithm corresponds to taking the limit of an infinite
ferromagnetic coupling ($J\to \infty $) in the
Ising Hamiltonian and that this isolates exactly the aging part of the response function.
With $d>1$ the $J \to \infty $ limit and the no bulk flip algorithm produce  
different dynamical evolutions and an argument analogous to the one in the $d=1$ case
cannot be made. What happens is that the limit $J \to \infty $ does not isolate $\chi _{ag}(t,t_w)$
because, besides freezing spins in the bulk, it also freezes most of the interfacial
spins. Notice that the no bulk flip dynamics does not obey detailed balance. This is simply due to
the fact that bulk spins are frozen. However, this is not a serious problem  
since we already know that by restoring moves in the bulk  
detailed balance is recovered producing  the stationary contribution in the
response function, which we are not interested in.

We have performed the simulations with the no bulk flip algorithm, after checking that the results are
consistent within 5 per cent with those of the global algorithms. In practice, we measure the 
quantity
\be
\chi_{ag}(t,t_w) = {1 \over {\cal N} h_0^2} \sum_{i=1}^{\cal N} \overline{\langle \sigma_i \rangle h_i}
\label{00}
\ee
where $h_i$ is a quenched configuration of an uncorrelated random field, which takes the
values $\pm h_0$ with probability $1/2$. The angular brakets stand for the average
over thermal histories, generated with the no bulk flip algorithm, and the overbar
denotes the average over random field configurations.
Simulations have been performed at $T/T_C=0.66$ 
for all values of $d$ (for the lattice size and the number of realizations see Table~\ref{tabella_chi}).
$\chi_{ag}(t,t_w)$ is measured in units $J^{-1}$ and time in units of Monte Carlo steps. For each
thermal history we have changed also the random field configuration.

\begin{center}
\begin{minipage}{6cm}
\begin{table}
\begin{tabular}{|c|*{6}{c|}}
    \hline       $t_w$   &
    \multicolumn{2}{c|}{d=2}&
    \multicolumn{2}{c|}{d=3}  &
    \multicolumn{2}{c|}{d=4} \\
    \hline          25   & 
    ${\cal N}$  &  realiz  &
    ${\cal N}$  &  realiz  &
    ${\cal N}$  &  realiz   \\
    \hline          25 & 
    $1024^2$  &  2000  &
    $100^3$   &  1000  &
    $42^4$    &  1600 \\
    \hline          50 & 
    $1024^2$  &  2000  &
    $150^3$   &  1000  &
    $68^4$    &  180 \\
    \hline    100      & 
    $1024^2$  &  2000  &
    $150^3$   &  1500  &
    $68^4$    &  75 \\
    \hline    250      & 
    $1024^2$  &  2300  &
    $150^3$   &  2700  &
              &   \\
    \hline     500      & 
    $1024^2$  &  15000  &
              &         &
              &   \\
    \hline    1000      & 
    $1024^2$  &  17000  &
              &         &
              &   \\
    \hline    1750      & 
    $1024^2$  &  17000  &
              &         &
              &   \\
    \hline    2500      & 
    $1024^2$  &  6000   &
              &         &
              &   \\
    \hline
\end{tabular}
\caption{Lattice size ${\cal N}$ and number of realizations in the computation of $\chi_{ag}(t,t_w)$ at different waiting times}
\label{tabella_chi}
\end{table}
\end{minipage}
\end{center}

First, we have obtained $a_{\chi}$ by plotting  $\chi_{ag}(t,t_w)$ versus $t_w$, for fixed values of
$x=t/t_w$.   In the range of $t_w$ explored there is excellent power law behavior. 
With $x=7$ we find (Fig.~\ref{fig.2})
$a_{\chi}=0.28$ for $d=2$, $a_{\chi}=0.47$ for $d=3$ and $a_{\chi}=0.50$ for $d=4$.
These numbers reproduce the results obtained 
previously~\cite{CLZ2001,Corberi2001,Corberi2003} confirming that $a_{\chi}$ in the Ising model
obeys closely Eq.s~(\ref{4.1}) and
(\ref{4.4}) with $\delta=1/2$,
$d_{\chi} =3$ and $d_L=1$ (see also Fig.~\ref{fig.1}). Furthermore, the observed behavior is with good accuracy
independent of $x$, as it is shown in the inset of Fig.~\ref{fig.2}.
The presence of a logarithmic correction at $d=3$ is hard to establish from
the data of Fig.~\ref{fig.2} since we have only one decade in $t_w$. In Ref.s~\cite{CLZ2001,Corberi2001}
where  $\chi_{ag}(t,t_w)$  was plotted against $t$ for fixed $t_w$
over four decades, the logarithmic behavior is accessible. Also, it should be mentioned
that  Eq.~(\ref{3.8}) is a phenomenological formula, so it is hard to say whether the measured
value $a_{\chi}=0.47$ for $d=3$ is due to logarithmic corrections or to some other effect not
captured by  Eq.~(\ref{3.8}). In any case, the quality of the data for $d=2$ allows to definitely rule
out $a_{\chi}=0.5$, predicted by the qualitative argument referred to in the Introduction and to be discussed shortly.  

\vspace{1cm}
\begin{figure}
\centerline{\psfig{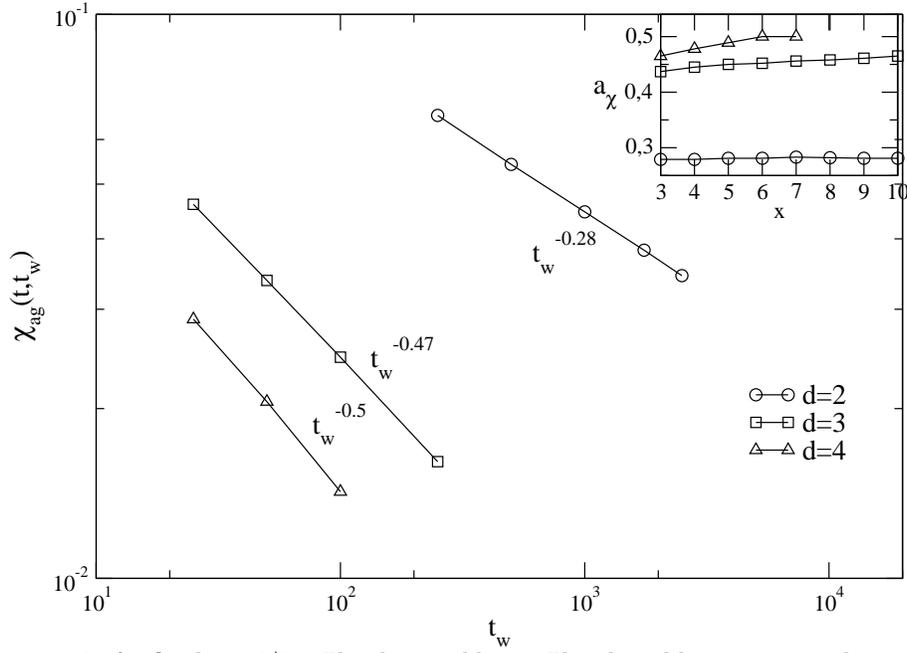}}
\caption{$\chi_{ag}(t,t_w)$ versus $t_w$ for fixed $x=t/t_w$. The slope yields
$a_{\chi}$. The plotted lines correspond to $x=7$. Values of $a_{\chi}$ for
different values of $x$ are depicted in the inset.
}
\label{fig.2}
\end{figure}

Next, in order to investigate the scaling function $\widehat{\chi}(x)$ in Eq. (\ref{3.4.3}),
notice that from Eq.s~(\ref{3.4.1}) and~(\ref{3.4.2}) follows the large $x$ behavior
\begin{eqnarray}
\widehat{\chi}(x) \sim   \left \{ \begin{array}{ll}
        x^{-a}   \qquad $for$ \qquad \alpha < 1 \\
        x^{-a}  \log x     \qquad $for$ \qquad  \alpha =1 \\
        x^{\alpha -a-1}   \qquad $for$ \qquad \alpha > 1.
        \end{array}
        \right .
        \label{5.5}
        \end{eqnarray}
Using the values of ${a_{\chi}}$ from Fig.~\ref{fig.2}, we have plotted 
$t_w^{a_{\chi}}\chi_{ag}(t,t_w)$ versus $x=t/t_w$ for different values of $t_w$ (Fig.s~\ref{fig.3},\ref{fig.4},\ref{fig.5}).

\vspace{1.5cm}
\begin{figure}
\centerline{\psfig{figure=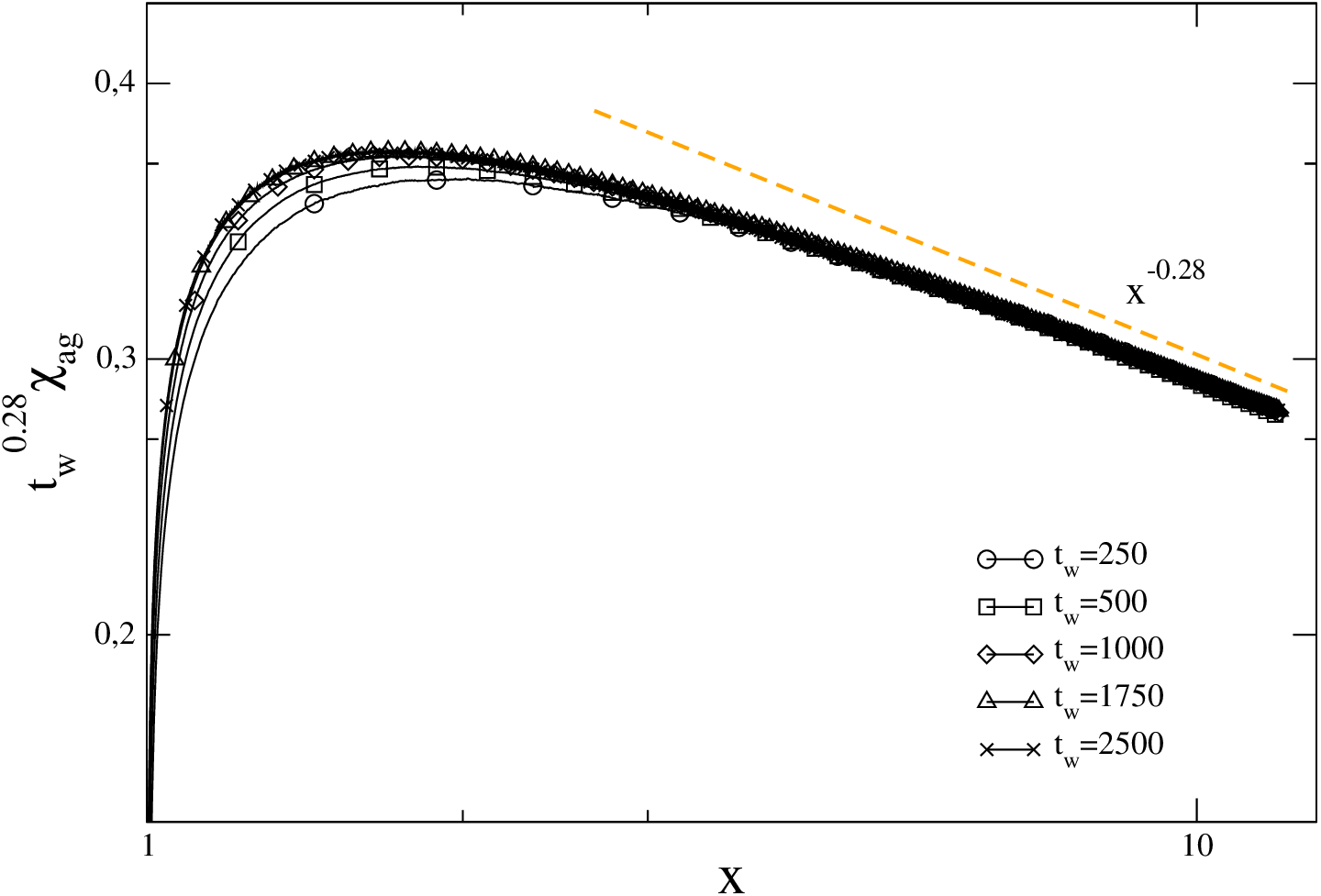,width=12cm,angle=0}}
\caption{Scaling function $\widehat{\chi}(x)$ for the $d=2$ Ising model with $T/T_C = 0.66$.}
\label{fig.3}
\end{figure}

\vspace{1cm}
\begin{figure}
\centerline{\psfig{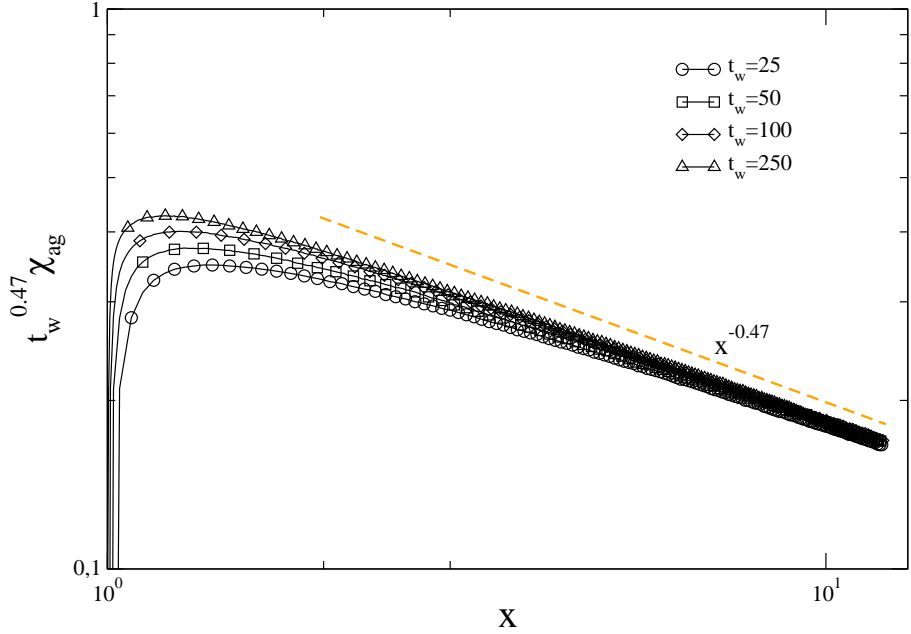}}
\caption{Scaling function $\widehat{\chi}(x)$ for the $d=3$ Ising model with $T/T_C = 0.66$.}
\label{fig.4}
\end{figure}
 
\vspace{1cm}
\begin{figure}
\centerline{\psfig{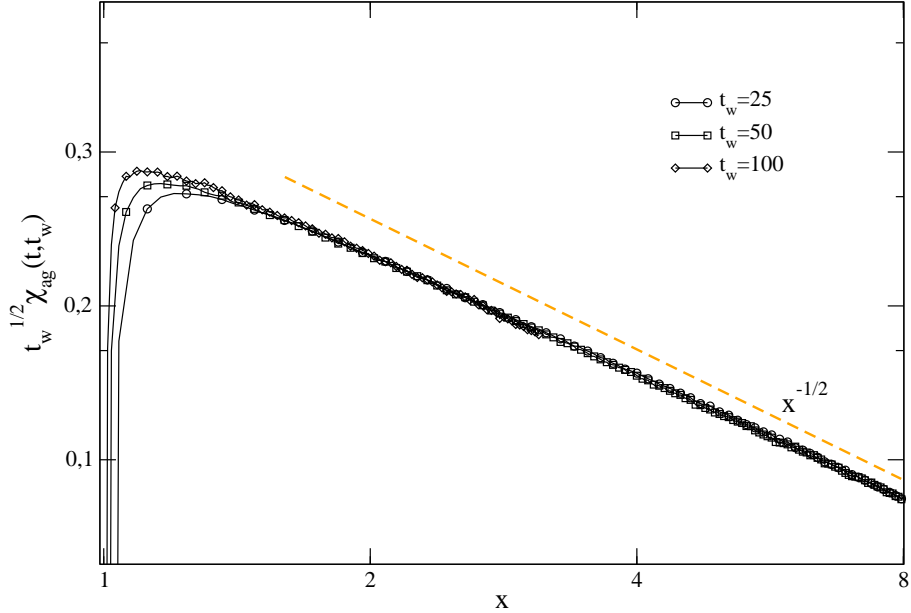}}
\caption{Scaling function $\widehat{\chi}(x)$ for the $d=4$ Ising model with $T/T_C = 0.66$.}
\label{fig.5}
\end{figure}
 
Collapse of the data is obtained 
for $x$ sufficiently large, where the scaling function decays with a power law
and an exponent which coincides with ${a_{\chi}}$.
This is consistent with Eq.~(\ref{5.5}) only if $\alpha =a +1/2$ as in
Eq.~(\ref{50.2}) and this rules out $\alpha =a +1$,  which ought to apply
according to the HPGL theory. Another way to see that $\alpha = a+1$
is untenable is that this
would imply that $\widehat{\chi}(x)$ goes to a constant for large $x$ when $\alpha >1$.
This, in turn, would lead to the unphisical conclusion that
$\chi_{ag}(t,t_w)$ does not decay to zero for large $t$ and
fixed $t_w$ when $d >d_{\chi}$. Therefore, we find that Eq.~(\ref{50.2}) holds for the Ising model not only
for $d=1$, but also at higher dimensionality.

In conclusion, Eq.s~(\ref{4.4}) and~(\ref{50.2}) are our main results for $a$ and $\alpha$ 
in the Ising model with $d$ ranging from $1$ to $4$ and with Eq.~(\ref{4.1}) 
explaining how $a$ is related to $a_{\chi}$.

\subsection{Qualitative conjecture on $a_{\chi}$}

We may now comment on the qualitative conjecture mentioned in the Introduction.
Stating~\cite{Barrat98,Franz98} that 
the aging contribution of ZFC ought to be proportional to
the density of defects and assuming scaling, one finds
\be
\chi_{ag}(t,t_w) \sim t_w^{-\delta} \widehat{\chi}(t/t_w) 
\label{3.6}
\ee
where $\delta$ is given by Eq.~(\ref{20.1}). This requires $a_{\chi} = \delta$ 
for all $d$ contrary to the evidence presented above and summarised 
in Eq.~(\ref{4.1}), which restricts the validity of  Eq.~(\ref{3.7}) to $d > d_{\chi}$.
This makes a big difference, for instance, in the $d=2$ Ising model where from Eq.~(\ref{4.1})
$a_{\chi} =1/4$, while from the above Eq.(\ref{3.6}) follows $a_{\chi} =1/2$. In order to understand why
Eq.~(\ref{3.6}) breaks down below $d_{\chi}$, let us go back to the behavior of $R_{ag}(t,s,t_0)$ in the
short time regime. From Eq.~(\ref{10.2}) we may write
\be
R_{ag}(t,s,t_0) \sim s^{-\delta} h(\tau,t_0)
\label{ph.1}
\ee
where $h(\tau,t_0)$ is some function of the time difference. The meaning of this is that the
response, due to an impulsive perturbation at the time $s$, is proportional to the density
of defects at that instant of time with a proportionality factor containing the
retardation effect. This does not hold anymore in the long time regme $\tau \gg s$. 
When the time interval $\tau$ is large with respect to $s$, multiple defect transits may have occurred 
through the observation
site, spoiling the form~(\ref{ph.1}). Sticking to the short time regime, i.e. taking
$t-t_w \ll t_w$ and using Eq.~(\ref{ph.1}), from the definition~(\ref{3.4}) follows
\be
\chi_{ag}(t,t_w) \sim t_w^{-\delta} \chi_s(t-t_w) 
\label{ph.2}
\ee
where $\chi_s(t-t_w)$ is a function of the time difference, which in Ref.s~\cite{Corberi2001,CLZ2001}
we have identified with the ZFC associated to a single defect. Now, Eq.s~(\ref{ph.2})
and~(\ref{3.6}) do require
\be
\widehat{\chi}(t/t_w) \sim \chi_s(t-t_w)
\label{ph.3}
\ee
which can hold only if both functions are constant. And this is precisely the point. As we have
explained in Ref.s~\cite{Corberi2001,CLZ2001} $\chi_s(t-t_w)$ contains the cumulative effect
on a single defect of the perturbation acting all along the time interval $(t_w,t)$. This
saturates rapidly to a constant when the defect degrees of freedom act paramagnetically and
the underlying defect motion is uncorrelated with the external field. However, at
dimensionalities low enough to reduce surface tension below the threashold where the external
field may take part in driving defect motion, $\chi_s(t-t_w)$ acquires a non trivial time dependence
which renders $a_{\chi} \neq \delta$ for $d < d_{\chi}$. Finally, notice that in the framework of the 
qualitative conjecture with $a_{\chi} = \delta$ independent of dimensionality, there is no 
explanation for the exact $d=1$ result $a_{\chi} = 0$. Instead, according to Eq.~(\ref{3.8})
this exact result, far from being an anomaly, is embedded as a limiting behavior in the smooth dimensionality 
dependence for $d < 3$.

\section{TRM} \label{trm}

Dealing with TRM, the separation~(\ref{3.2}) gives $\rho(t,s)= \int_0^{t_w} R_{st}(t-s)
+ \int_0^{t_w} R_{ag}(t,s)$. Contrary to what happens for ZFC, where $\chi_{st}$ for long time saturates 
to a constant, here for the stationary contribution there are two possibilities:
i) if $R_{st}(t-s)$ decays exponentially $\rho_{st}(t-t_w)$ also decays exponentially
or ii) if $R_{st}(t-s)$ decays with a power law, like in the large $N$ model,
$\rho_{st}(t-t_w)$ is subdominant with respect to $\rho_{ag}(t,t_w)$. In both cases we can neglect $\rho_{st}$ 
and with it the distinction between  $\rho$ and $\rho_{ag}$.

As mentioned previously, TRM is affected by preasymptotic contributions which cannot be eliminated.
This makes it quite difficult to establish if the asymptotic behavior has been reached in the simulations
and ultimately to have a reliable estimate of $a$. In order to unravel what is
the effect of the preasymptotic contributions on the scaling behavior of TRM, we have resorted as a guide
to the solution of the large $N$ model (section~\ref{larg}). Here, we anticipate the results.

Assuming $t_w > t_{sc}$, in the large $N$ case there exists a dimensionality $d_{\rho}=4$ such that 
for $ d < d_{\rho} $ TRM undergoes a crossover with a characteristic time $t^*$, which may also
be much larger than $t_{sc}$. Introducing the effective exponent
\be
a_{\rho,eff}  =  \left. - \frac{\partial \log \rho(t,t_w,t^*)}{\partial \log t_w} \right|_{t/t_w} 
\label{2.34.2bis}
\ee
one finds
\be
a_{\rho,eff} =    \left \{ \begin{array}{ll}
        \lambda/z \qquad $for$ \qquad t_w  \ll t^*  \\
        a  \qquad $for$ \qquad t_w  \gg t^*. 
        \end{array}
        \right .
        \label{2.34.20}
\ee

For $d  = d_{\rho}$ there is a crossover from a pure power law to a power law with
logarithmic correction
\be
  \rho(t,t_w,t^*)   = \left \{ \begin{array}{ll}
        t_w^{-\lambda/z} E(t/t_w)    \qquad $for$ \qquad t_w  \ll t^*  \\
        t_w^{-\lambda/z} \log(t_w/t_{sc})  E(t/t_w)   \qquad $for$ \qquad t_w  \gg t^*. 
        \end{array}
        \right .
        \label{2.34.3}
        \ee

Finally, for $d > d_{\rho}$ one has the simple power law
\be
  \rho(t,t_w,t^*)   =t_w^{-\lambda/z} E(t/t_w)
\label{2.34.4bis}
\ee
and for all values of $d$, in the time regime considered the scaling function obeys
\be
E(x) \sim x^{-\lambda/z}.
\label{2.34.40bis}
\ee

\vspace{5mm}

Taking this pattern as a guide (with $d_{\rho}$, $t^*$ and exponents model dependent)
let us now turn to simulations of the Ising model. Analysing data, the first thing to do is to check if 
a behavior of the type
\be
\rho(t,t_w) \sim t_w^{-a_{\rho}}\widehat{\rho}(t/t_w)
\label{4.100}
\ee
holds. If this is the case and if an exponent $a_{\rho}$ can be meaningfully extracted,
the next problem is relating  $a_{\rho}$ to $a$. According to the  
behavior found in the large $N$ model, the identification $a_{\rho} = a$ can be made
only if $d < d_{\rho}$ and $t_w \gg t^*$. Numerical results for TRM in the Glauber-Ising
model have been first obtained by HPGL~\cite{Henkel2001}. Plotting $\rho(t,t_w)$
against  $x=t/t_w$ for different $t_w$ in the range $t_w \in (25, 250)$ for $d=2$ and
$t_w \in (15, 100)$ for $d=3$ they have obtained for $a_{\rho}$ a
result of the form
\be
    a_{\rho} =\left \{ \begin{array}{ll}
        1/2   \qquad $with log corrections for$ \qquad d=2 \\
        1/2   \qquad $for$ \qquad d = 3 
        \end{array}
        \right .
        \label{3.10}
        \ee
and they have made the identification $a=a_{\rho}$.

The next round of simulations has been carried out by us~\cite{Corberi2003}  at the same temperatures
and for the same system size as HPGL, but extending the range of $t_w$
up to $2500$ for $d=2$ and $250$ for $d=3$. Performing a different
data analysis, i.e.
plotting $\rho(t,t_w)$ versus $t_w$ for fixed $x=t/t_w$, we have found
good agreement between the slope of the curves in the large $t_w$ region, which in the log-log plot gives the
effective exponent (\ref{2.34.2bis}), and the known values of $\lambda/z$ for the
Ising model
($\lambda/z=5/8$ for $d=2$ and  $\lambda/z=3/4$ for $d=3$). This is good evidence
that in the scalar case TRM follows the crossover pattern obtained in
the large $N$ model when $d < d_{\rho}$ and with a crossover time $t^*$ larger than the
maximum $t_w$ that we have reached in the simulations. Furthermore, on the basis 
of our data, we have
estimated that the largest $t_w$ used by HPGL in Ref.~\cite{Henkel2001}  was not enough to enter 
the scaling regime (i.e.  they had always $t_w \leq t_{sc}$) and therefore the values of $a_{\rho}$
they have obtained do not warrant any statement neither on the asymptotic
value of  $a_{\rho}$ nor on $a$. 
Our longer range of $t_w$ seems to be barely sufficient to 
enter the preasymptotic region where $a_{\rho,eff} = \lambda/z$, 
suggesting that both $d=2$ and $d=3$ are smaller than  $d_{\rho}$, whose
value in the Ising model, so far, we do not know. Hence, in order to observe the
asymptotic exponent one shoud go to much longer waiting times $t_w$.

Henkel and Pleimling~\cite{Henkel2003} have produced new simulations for $d=2$
extending the range of $t_w$ up to $5000$. Plotting
$\rho(t,t_w)$ versus $t_w$ for fixed $x$ and adhering to the point
of view that the TRM data are affected by a long crossover, they claim i) to have
succeded in going past the crossover time reaching the asymptotic region
and ii) to have found that Eq.~(\ref{3.10}) is verified.
The objection to this claim is that in $d=2$ one has 
$\lambda/z = 5/8 > 1/2  > a_{\chi} =1/4$. Therefore, even if 
a decrease of the slope from a number close to $\lambda/z=0.625$ toward $0.5$ 
is observed over a narrow time window, there is no way
to decide whether the true asymptotic regime has been reached or  
the slope might still keep on decreasing, by going further with $t_w$, 
untill reaching asymptotics at $0.25$.

In other words, the new simulations in Ref.~\cite{Henkel2003} leave the issue undecided and yet
longer simulations are needed. Despite, by now, there is sufficient evidence
that TRM is not the most efficient and reliable way to get to the
exponent $a$, we have undertaken new simulations with $t_w$ up to $5000$ for $d=2$
and $500$ for $d=3$ (for the lattice size and number of realizations see Table~\ref{tabella_rho}). 

\begin{center}
\begin{minipage}{6cm}
\begin{table}
\begin{tabular}{|c|*{4}{c|}}
    \hline       $t_w$   &
    \multicolumn{2}{c|}{d=2}&
    \multicolumn{2}{c|}{d=3}  \\
    \hline          25   & 
    ${\cal N}$  &  realiz  &
    ${\cal N}$  &  realiz   \\
    \hline          25 & 
    $1024^2$  &  2000  &
    $100^3$   &  1500  \\
    \hline          50 & 
    $1024^2$  &  2000  &
    $150^3$   &  2500  \\
    \hline    100      & 
    $1024^2$  &  2000  &
    $150^3$   &  2500  \\
    \hline    250      & 
    $1024^2$  &  13000  &
    $150^3$   &  2500  \\
    \hline     500      & 
    $1024^2$  &  16000  &
     $160^3$  &   2500 \\
    \hline    1000      & 
    $1024^2$  &  18000  &
              &   \\
    \hline    1750      & 
    $1024^2$  &  23000  &
              &   \\
    \hline    2500      & 
    $1024^2$  &  13000   &
              &   \\
    \hline    5000      & 
    $2048^2$  &  7000   &
              &   \\
    \hline
\end{tabular}
\caption{Lattice size ${\cal N}$ and number of realizations in the computation of $\rho_{ag}(t,t_w)$ at different waiting times}
\label{tabella_rho}
\end{table}
\end{minipage}
\end{center}

The double logarithmic plot of $\rho(t,t_w)$ versus $t_w$ for
fixed $x$ shows (Fig.~\ref{fig.6}) that a power law behavior, possibly, sets in only in the
region of the largest $t_w$ reached. Taking the slope in this region as a measure of 
$a_{eff,\rho}$ we find values (inset of Fig.~\ref{fig.6}) which lie above $0.5$ for all $x$
and that are just below $\lambda/z=0.625$ for $d=2$ and $\lambda/z=0.75$ for $d=3$.
Hence, although we have reached the same maximum value of $t_w$ as in 
Ref.~\cite{Henkel2003} for $d=2$ and we have gone much farther for $d=3$, we may state
that no evidence of asymptotic behavior with $a_{\rho}= 1/2$ is found. Rather, the
combination of the $d=2$ and $d=3$ data in Fig.~\ref{fig.6} shows unequivocally that, at best, only the
onset of the scaling region is entered where $a_{\rho,eff}$ is about to take 
the preasymptotic value $\lambda/z$, confirming the picture obtained in our 
previous work~\cite{Corberi2003}.

\vspace{1cm}
\begin{figure}
\centerline{\psfig{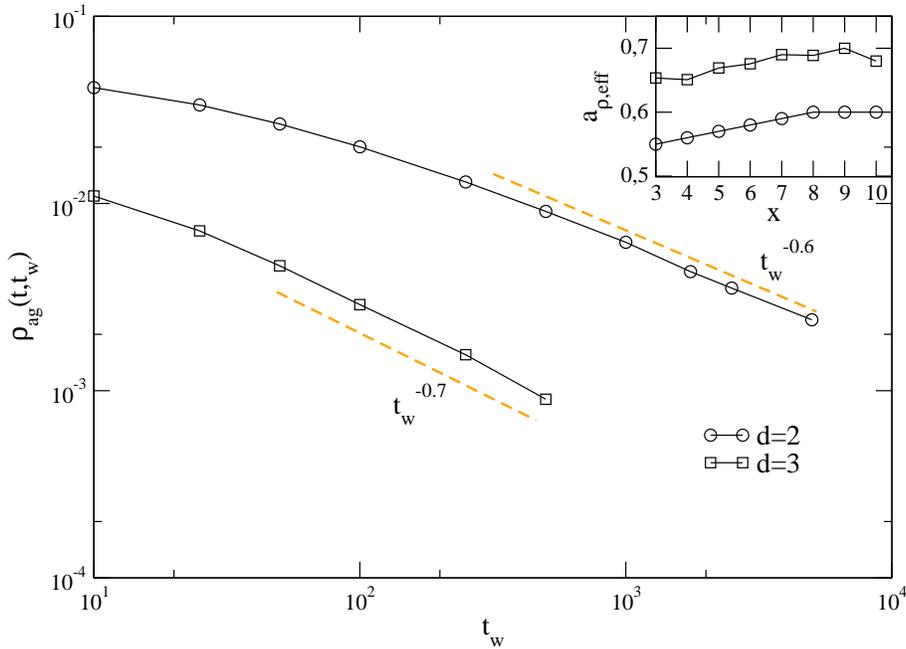}}
\caption{$\rho_{ag}(t,t_w)$ versus $t_w$ for fixed $x=7$. The slopes
in the large $t_w$ region yielding $a_{\rho,eff}$ for different 
values of $x$ are depicted in the inset.}
\label{fig.6}
\end{figure}

In summary, we have accumulated sufficient numerical evidence to establish that
TRM data fit in the general pattern of behavior abtained from the solution
of the large $N$ model, with $d_{\rho} > 3 $ and a value of $t^*$ which is greater than
the largest  $t_w$ reached so far. Therefore, since asymptotics has not been
reached, no statement on $a$ can be made from the present knowledge of
TRM.

Finally, let us make a comment on the quotation in 
Ref.s~\cite{Henkel2001,HPM2003,HM2003} of the analytical solution of the GAF
approximation by Berthier et al.~\cite{Berthier} as a support to the claim that $a$ is
given by Eq.~(\ref{3.10}). In fact, here is where is most evident the type of
confusion that can be made by not being careful about which exponent one is
talking about.
In their computation Berthier et al. find $a_{\chi} = 1/2$ for $d \geq 2$ with
logarithmic correction at $d=2$, as in Eq.~(\ref{3.10})which, however, is meant for $a$. What one
should have clear in mind is that they  
compute an $a_{\chi}$ for $d \geq d_{\chi}$, i.e. right where $a_{\chi} \neq a$.
This can be checked recalling that in the GAF approximation $a$ is given by
Eq.~(\ref{2.13}) and that $d_{\chi}=2$.
Hence, for $d=2$ the logarithmic correction belongs to  $a_{\chi}$ and
not to $a$. For $d=3$ it is $a_{\chi}$ that takes the value $1/2$, while from Eq.~(\ref{2.13}) 
follows $a=1$.
So, the results of Berthier et al. certainly cannot be quoted if one wants to 
identify with $a$ an exponent obeying Eq.~(\ref{3.10}).

\section{TRM and ZFC in the large N model} \label{larg}

In this section we study in detail the large $N$ model~\cite{Corberi2002,Mazenko} as a useful 
example which gives the complete picture of what happens 
when looking at the different response functions introduced above.

Consider a system with vector order parameter $\vec \phi(\vec
x)=(\phi_1(\vec x),...,\phi_N(\vec x))$ and Hamiltonian of the
Ginzburg-Landau form
\be
    {\cal H} [\vec \phi] = \int  d ^d x \left [ \frac{1}{2} ( \nabla
    \vec \phi )^2 + \frac{r}{2} \vec \phi ^2 + \frac{g}{4N}(\vec \phi
    ^2)^2 \right ]
    \ee
where $r<0$, $g>0$. In the large $N$ limit the equation of motion
for the generic component of the order parameter in Fourier space
is given by \cite{Corberi2002}
\be
    \frac{\partial  \phi (\vec k ,t) }{\partial t} = - [k^2 + I(t)
    ] \phi( \vec k,t) +  \eta (\vec k,t)
    \label{6.2}
    \ee
where $\eta(\vec k,t)$ is a gaussian white noise with expectations
\be
      \begin{array}{ll}
      \langle \vec \eta (\vec k, t) \rangle = 0  \\
      \langle \eta  (\vec k,t) \eta  (\vec k',t') \rangle = 2T
      (2\pi)^d
       \delta(\vec k +\vec k') \delta(t-t')
                       \end{array}
    \label{6.3}
    \ee
$T$ is the temperature of the quench and the function of time
\be
    I(t) = r + g \langle \phi^2(\vec x,t)\rangle
    \label{6.4}
    \ee
must be determined self-consistently, with the average on the
right hand side taken both over thermal noise and initial
conditions. The formal solution of (\ref{6.2}) is given by
\be
     \phi (\vec k,t) = R(\vec k,t,0) \phi _0(\vec k)+ \int _0
    ^t ds R(\vec k,t,s) \eta (\vec k,s)
    \label{6.5}
    \ee
where $R(\vec k, t,s)$ is the response function
\be
    R(\vec k, t,s)=\frac{Y(s)}{Y(t)}e^{-k^2(t-s)}
    \label{6.6}
    \ee
with $Y(t)=\exp[Q(t)]$ and $Q(t)= \int _0 ^t ds I(s)$. With an
uncorrelated initial state at very high temperature the initial
condition $\phi_0(\vec k)=\phi(\vec k,t=0)$ can be taken to be
gaussianly distributed with expectations
\be
      \begin{array}{ll}
      \langle \phi _{0} (\vec k)\rangle  = 0  \\
       \langle \phi _{0}(\vec k)
       \phi _{0}(\vec k')\rangle  = \Delta (2\pi)^d
        \delta(\vec k+\vec k') .
                       \end{array}
    \label{initial2}
    \ee
The actual solution is obtained once the function $Y(t)$ is
determined. In order to do this notice that from the definition
of $Y(t)$ follows
\be
    \frac{dY^2(t)}{dt}=2[r+g\langle \phi^2(\vec x,t \rangle ]Y^2(t).
    \label{6.8}
    \ee
Writing $\langle \phi^2(\vec x,t) \rangle $ in terms of the structure
factor
\be
    \langle \phi^2(\vec x,t) \rangle =\int 
\frac{d^d k}{(2\pi)^d}C(\vec k,t) e^{-k^2/\Lambda^2}
    \label{6.9}
    \ee
where $\Lambda$ is the momentum cutoff and using Eq.(\ref{6.5})
to evaluate $ \langle \phi(\vec k,t)\phi(\vec k',t) \rangle =C(\vec
k,t)(2\pi)^d\delta(\vec k+\vec k') $ we obtain

\be
    C(\vec k,t) =R^2(\vec k,t,0) \Delta + 2T \int _0^t
    ds R^2(\vec k,t,s) .
    \label{6.10}
    \ee
Then, inserting Eq.~(\ref{6.9}) into Eq.~(\ref{6.8}) we obtain the 
integro-differential equation
\be
    \frac{dY^2(t)}{dt}=2rY^2(t)+2g\Delta f\left (t+\frac{1}{2\Lambda
    ^2}\right ) +4gT \int _0 ^t ds f\left (t-s+\frac{1}{2\Lambda
    ^2}\right ) Y^2(s)
    \label{6.11}
     \ee
where $ f(x)\equiv \int \frac{d^dk}{(2\pi )^d}e^{-2k^2x}=(8\pi
x)^{-\frac{d}{2}}$. After solving this equation, 
the response function is given by
\be
    R(t,s,t_0)=\int\frac{d^dk}{(2\pi)^d}R(\vec k,t,s) e^{-k^2/\Lambda^2}
    =(4\pi)^{-d/2}\frac{Y(s)}{Y(t)}(t-s+t_0)^{-d/2}
    \label{6.12}
    \ee
where $t_0=1/(2\Lambda^2)$. 

Let us now come to the identification of the general structure of
Eq.~(\ref{1.2}). Since in the
stationary regime $Y(t)$ is time independent, we immediately
obtain
\be
    R_{st}(t-s,t_0)=(4\pi)^{-d/2}(t-s+t_0)^{-d/2}
    \label{2.12.1}
    \ee
and
\be
    R_{ag}(t,s,t_0)=(4\pi)^{-d/2}\left[\frac{Y(s)}{Y(t)}-1\right](t-s+t_0)^{-d/2}
    .
    \label{2.12.2}
    \ee
Notice that $R_{st}(t-s,t_0)$ is temperature independent, implying that there is a stationary
response also at $T=0$. This holds for soft spins, while for Ising spins there is no stationary
response at $T=0$.

Next, in order to investigate the scaling properties 
we must first learn about the time dependence of $Y(t)$. 
We do this in the $T=0$ case, since quenches below the
critical point are controlled by the $T=0$ fixed point \cite{Bray94}. 
Making the ansatz
$Y(t)=At^{-\omega}$ from (\ref{6.11}) one gets
\be
    A \omega t^{-(2\omega +1)}=r A t^{-2\omega}+\frac{2g
    \Delta}{(8\pi)^{d/2}}(t+t_0)^{-d/2}
    \label{6.13}
    \ee
and assuming that the left hand side is negligible one
finds $\omega=d/4$ with $A=(8\pi)^{-d/2}\Delta/M_0^2$, where
$M_0=\sqrt{-r/g}$ is the zero temperature magnetization. 
This is consistent if, in addition to $t \gg t_0$, one has also
\be
    t \gg t_{sc}=-d/(4r)
    \label{6.14}\
    \ee
where the characteristic time $t_{sc}$ sets the time scale 
over which the three terms in
Eq.(\ref{6.13}) are all of the same order of magnitude.
Therefore, $t_{sc}$ is the characteristic time separating the early
from the late stage.

The above described behavior of $Y(t)$ is illustrated in
Fig.~\ref{fig.7} displaying the numerical solution of
Eq.~(\ref{6.11}) for different values of $r$. In all numerical computations we will
take $\Delta =1$, $T=0$ and time is measured in units $t_0$. The onset of the scaling behavior
is sharp and we have identified $t_{sc}$ with the time where the power law begins
(inset of Fig.~\ref{fig.7}). Then, for $s,t > t_{sc}$ 
from Eq.s~(\ref{6.12}) and (\ref{2.12.2}) we have
\be
R(t,s,t_0) = s^{-(1+a)}\widetilde{f}(t/s,t_0/s)
\label{6.15}
\ee
with
\be
\widetilde{f}(x,y) = (4\pi)^{-d/2} x^{\omega} (x-1+y)^{-(1+a)}
\label{6.16}
\ee
where
\be
a= (d-2)/2.
\label{2.18.1}
\ee
as in Eq. (\ref{2.10}).

The connection between $\omega$ and $\lambda/z$ can be established from the
autocorrelation function. Keeping on considering $T=0$, from
$C(\vec{k},t,s) = R(\vec{k},t,0)R(\vec{k},s,0)\Delta$ follows
\begin{eqnarray}
    C(t,s,t_0) & = & \int\frac{d^dk}{(2\pi)^d}C(\vec k,t,s) e^{-k^2/\Lambda^2} \\
    & = & (4\pi)^{-d/2}\Delta s^{2\omega-(1+a)}  (t/s)^\omega
    [t/s+1+t_0/s]^{-(1+a)}.
    \label{2.18.5}
\end{eqnarray}
The requirement $\lim_{t \rightarrow \infty} C(t,t) = M^2_0$ implies
\be
2\omega = 1+a
\label{2.18.6}
\ee
and comparing Eq.~(\ref{2.18.5}) with  Eq.~(\ref{1.2}) we find
\be
\omega = \lambda/z.
\label{2.18.7}
\ee
Hence, in the large $N$ model, $\lambda$ and $a$ are not independent
exponents, since from Eq.s~(\ref{2.18.6}) and (\ref{2.18.7}) follows
\be
\lambda= 1+a.
\label{2.18.8}
\ee
Nonetheless, for generality we shall keep on using the notation
with two different exponents  $\lambda$ and $a$.

\vspace{1cm}
\begin{figure}
\centerline{\psfig{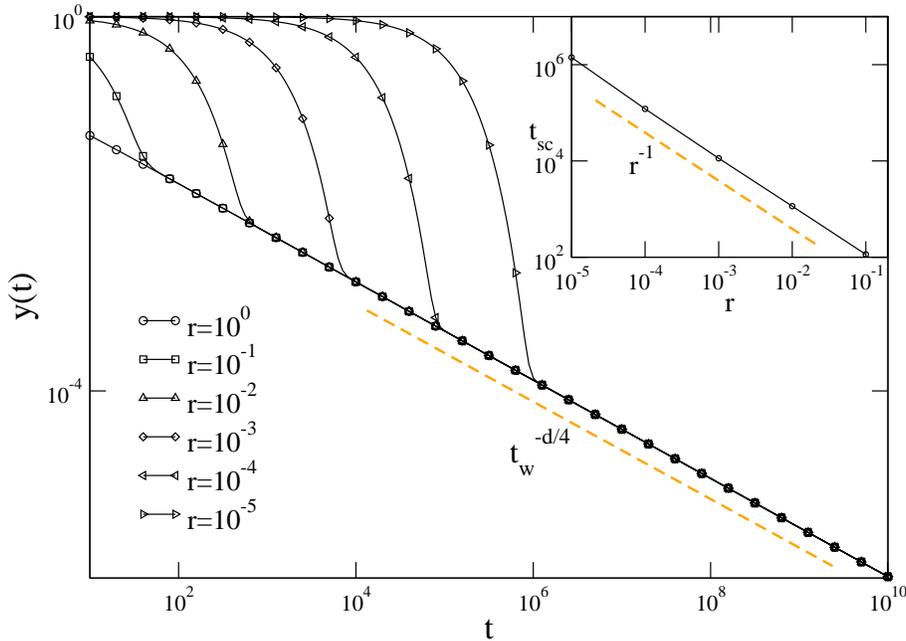}}
\caption{$Y(t)$ for different values of $r$ and $T=0$. $t_{sc}$ is estimated at the onset
of the power law behavior and plotted against $r$ in the inset.
}
\label{fig.7}
\end{figure}

Finally, for the aging contribution (\ref{2.12.2}) we may write
\be
R_{ag}(t,s,t_0)=s^{-(1+a)}f(t/s,t_0/s)
\label{6.17}
\ee
with
\be
f(x,y) = (4\pi)^{-d/2}( x^{\omega} -1) (x-1+y)^{-(1+a)}
\label{6.18}
\ee
and writing $\omega=1+a-\lambda/z$ Eq.~(\ref{2.9}) is recovered. 

The above result shows that
in the large $N$ model it is not only $R_{ag}(t,s)$ to scale,
but also the full autoresponse function $R(t,s)$. This, obviously,
means that  $R_{st}(t-s)$ obeys scaling, as it can be checked 
immediately from Eq.~(\ref{2.12.1}) and this is a consequence of the
fact that the whole low temperature
phase is critical.

\subsection{TRM}

We now explore the properties of the IRF in the large $N$ model. 
Let us begin from TRM. Since the explicit forms~(\ref{2.12.1}) and~(\ref{6.17}) with~(\ref{6.18}) 
show that $R_{st}(t-s,t_0)$ decays faster than  $R_{ag}(t,s,t_0)$ with the time
separation $t-s$, taking $t \gg t_w$ and using the definitions~(\ref{3.2}) and~(\ref{3.3})
the stationary contribution to TRM can be neglected. Hence, in the following
we will ignore the distinction between $\rho$ and $\rho_{ag}$. Furthermore, taking
$t_w > t_{sc}$ and dropping the dependence on $t_0$ we can write
\be
\rho(t,t_w,t_{sc}) = \frac {t^{-\lambda/z}}{(4\pi)^{d/2}A} \left [ \int_0^{t_{sc}}ds Y(s) +
\int_{t_{sc}}^{t_w}ds Y(s) \right ]
\label{2.17.0}
\ee
where we have separated the preasymptotic from the asymptotic contribution in
the integral. We shall see shortly that the first one plays a crucial role.
Introducing the notation $B(t_{sc})=\int_0^{t_{sc}}ds Y(s)$ and  using
$Y(s) = A s^{-\lambda/z-(a+1)}$ in the second integral, we find 
\be
\rho(t,t_w,t_{sc}) =   t_w^{-\lambda/z} \left [ K_0 + K_1 t_{w}^{\lambda/z-a}\right ]
(t/t_w)^{-\lambda/z}
\label{2.18.0}
\ee
where
\be
K_0 = (4\pi)^{-d/2} \left [\frac{B(t_{sc})}{A} - \frac{t_{sc}^{\lambda/z-a}}{(\lambda/z-a)} \right ]
\label{6.20} 
\ee
and
\be
K_1 = \frac{1}{(4\pi)^{d/2}   (\lambda/z-a)}.
\label{6.21} 
\ee
Eq.~(\ref{2.18.0}) is the main result from which follows the non trivial
dependence of $a_{\rho}$ on dimensionality.
Notice, that all the dependence on the preasymptotic behavior is collected
in $K_0$ and the very presence of this non vanishing term entails that
the asymptotic power governing TRM is either $\lambda/z$ or $a$ according
to the sign of $(\lambda/z-a)$. Therefore, writing $\lambda/z -a = (d_{\rho}-d)/
d_{\rho}$ with $d_{\rho} =4$ we have a crossover for $ d < d_{\rho}$,
logarithmic corrections for $d = d_{\rho}$ and a correction to scaling 
for $d > d_{\rho}$.

Introducing the characteristic time
\be
t^* = \left (\frac {K_0 }{|K_1|} \right )^{\frac{1}{\lambda/z-a}} 
\label{6.22} 
\ee  
Eq.~(\ref{2.18.0}) for  $d \neq d_{\rho}$ can be rewritten as
\be
\rho(t,t_w,t^*) =  t_w^{-\lambda/z} \widetilde{E}(t/t_w,t^*/t_w)
\label{6.35} 
\ee
with
\be
\widetilde{E}(x,y) =  K_0 \left [ 1 \pm y^{\lambda/z-a} \right ]x^{-\lambda/z}
\label{2.35.1} 
\ee
where  the $+$ and  $-$ signs apply to $d < d_{\rho}$ and $ d> d_{\rho}$,
respectively. In the first case the crossover time $t^*$ is given by (Appendix)
\be
t^*/t_{sc} \sim t_{sc}^{d/(4-d)}
\label{6.29} 
\ee
showing that $t^*$ is a new time scale which can become much larger than $t_{sc}$.
Instead, in the second case from  Eq.~(\ref{6.22})
follows (Appendix)
\be
t^*/t_{sc}  < 1
\label{6.30} 
\ee
implying $t_w/t^* > 1$ for any $t_w > t_{sc}$.
Finally, for  $d = d_{\rho}$ from Eq.~(\ref{2.18.0}) we have
\be
\rho(t,t_w,t^*,t_{sc}) =  (4\pi)^{-d/2} t_w^{-\lambda/z} \left [1 + \frac{\log (t_w / t_{sc})}{\log (t^*/ t_{sc})}
\right ] \log (t^*/ t_{sc}) (t/t_w)^{-\lambda/z}
\label{6.32} 
\ee
where $t^*$ is given by
\be
t^*/t_{sc} = e^{Ct_{sc}}
\label{6.34} 
\ee
and $C$ is a constant (Appendix).

Therefore, as anticipated in section~\ref{trm}, the scaling properties of TRM exhibit the
following dimensionality dependence

\vspace{5mm}

$ {\bf d < d_{\rho}} $

\vspace{3mm} 

There is a crossover with the effective
exponent
\begin{eqnarray}
a_{\rho,eff} & = &  - \left. \frac{\partial \log \rho(t,t_w,t^*)}{\partial \log t_w} \right|_{t/t_w} 
\nonumber\\
& = &\lambda/z  - \left [ \frac{(t_w/t^*)^{\lambda/z-a}} 
  {1+ (t_w/t^*)^{\lambda/z-a} } \right ](\lambda/z-a) 
        \label{2.34.2}
        \end{eqnarray}
yielding
\be
a_{\rho,eff} =    \left \{ \begin{array}{ll}
        \lambda/z \qquad $for$ \qquad t_w  \ll t^*  \\
        a  \qquad $for$ \qquad t_w  \gg t^*. 
        \end{array}
        \right .
        \label{pppppp}
\ee

\vspace{5mm}

${\bf d  = d_{\rho}}$

\vspace{3mm} 

The crossover involves a logarithmic correction
\be
  \rho(t,t_w,t^*)   = \left \{ \begin{array}{ll}
        t_w^{-\lambda/z} E(t/t_w)    \qquad $for$ \qquad t_w  \ll t^*  \\
        t_w^{-\lambda/z} \log(t_w/t_{sc})  E(t/t_w)   \qquad $for$ \qquad t_w  \gg t^*. 
        \end{array}
        \right .
        \label{2.34.3bbis}
        \ee

\vspace{5mm}

${\bf d > d_{\rho}}$

\vspace{3mm}

There is a pure power law for all $t_w >t_{sc} $
\be
  \rho(t,t_w,t^*)   =t_w^{-\lambda/z} E(t/t_w)
\label{2.34.4}
\ee
with
\be
E(x) \sim x^{-\lambda/z}.
\label{2.34.40}
\ee 

In the end, in the large $N$ model the relation between $a$ and the exponent 
$a_{\rho}$ appearing in Eq.~(\ref{4.100}) is given by 
\be
    a_{\rho} = \left \{ \begin{array}{ll}
        a  \qquad $for$ \qquad d < d_{\rho}  \\
        \lambda/z  \qquad $with log corrections for$ \qquad d=d_{\rho} \\
	\lambda/z   \qquad $for$ \qquad d > d_{\rho} 
        \end{array}
        \right .
        \label{4.150}
        \ee
where $d_{\rho} =4$.

In order to illustrate the behavior of TRM we have solved numerically
for $\rho(t,t_w)$. In Fig.~\ref{fig.8}  we have plotted the effective exponent 
(\ref{2.34.2}) versus $t_w$ for
different values of $r$ (giving rise to different values of $t_{sc}$),
with fixed $x=t/t_w=20$ and for $d=2.1 < d_{\rho}$. The curves show quite
clearly three different regimes: the early regime to the left of the
peak followed by the intermediate regime going like  $t_w^{-\lambda/z}$,
whose size depends on $t_{sc}$, and eventually by the late stage regime going
like  $t_w^{-a}$. The value $t_{max}$ of
$t_w$ at the peak can be identified with $t_{sc}$ since it depends on $r$
according to Eq.~(\ref{6.14}) (see inset of Fig.~\ref{fig.8}). For completeness
we have plotted the same figure for $d=5 > d_{\rho}$ (Fig.~\ref{fig.9}) which shows the 
existence only of the early regime followed immediately by the asymptotic
regime with the exponent $\lambda/z$ (without any crossover or intermediate 
scaling regime) according to Eq.~(\ref{2.34.4}).

\vspace{1cm}
\begin{figure}
\centerline{\psfig{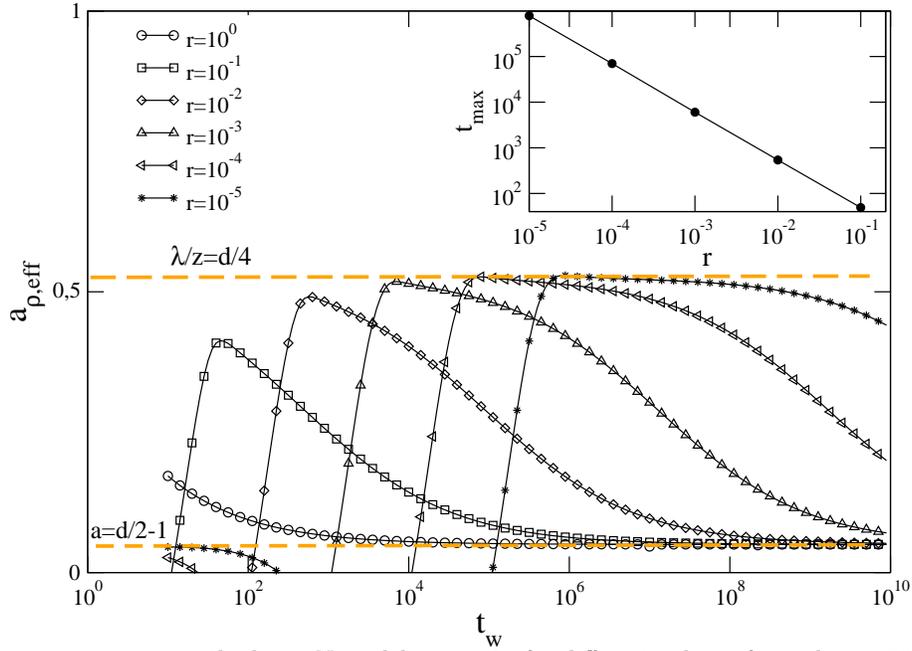}}
\caption{Effective exponent $a_{\rho,eff}$ in the large $N$ model 
versus $t_w$ for different values of $r$
with $x= 20$, $d=2.1$ and $T=0$. The value of $t_w$ at the maximum corresponds to
$t_{sc}$ as shown in the inset.
}
\label{fig.8}
\end{figure}

\vspace{1cm}
\begin{figure}
\centerline{\psfig{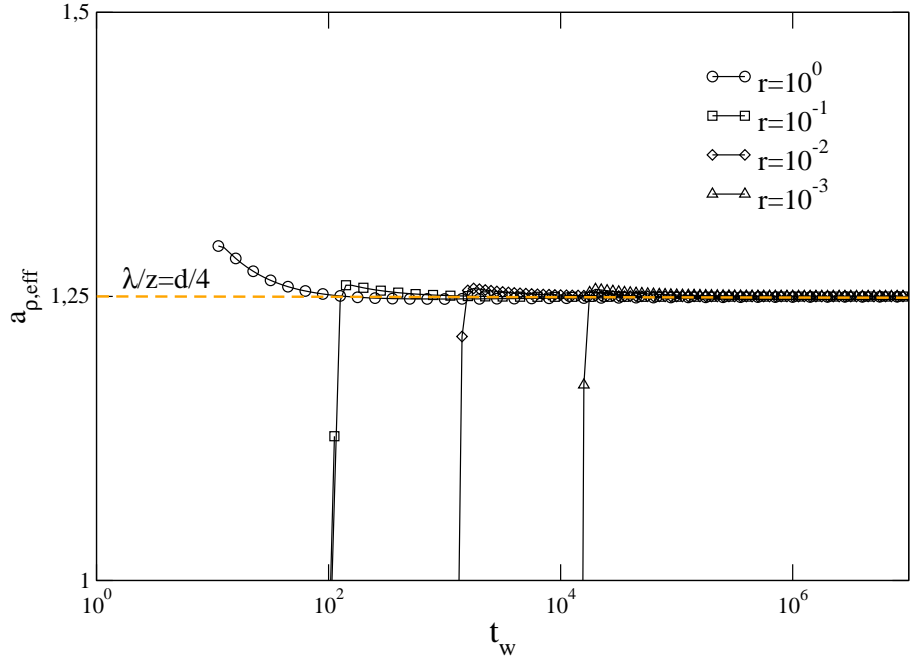}}
\caption{Effective exponent $a_{\rho,eff}$ in the large $N$ model 
versus $t_w$ for different values of $r$
with $x= 20$, $d=5$ and $T=0$.}
\label{fig.9}
\end{figure}

\subsection{ZFC}

Taking $t_w > t_{sc}$ and using the definitions (\ref{3.2}), (\ref{2.12.1}), 
(\ref{2.12.2}) we have
\be
\chi_{st}(t-t_w,t_0) = \frac{2t_0^{1-d/2}}{ (4\pi)^{d/2}(d-2)}  
\left \{ 1 - [(t-t_w)/t_0 + 1 ]^{-a} \right \}
\label{7.1}
\ee
and
\be
\chi_{ag}(t,t_w,t_0) = t_w^{-a} F(t/t_w,t_0/t_w)
\label{7.2}
\ee
with
\be
F(x,y) = (4\pi)^{-d/2} x^{-a} \int_{1/x}^1 du 
(u^{\lambda/z - (1+a)} -1) (1-u +y/x)^{-(1+a)}.
\label{7.3}
\ee
Therefore, in order to establish how $\chi_{ag}$ scales with $t_w$ it is necessary
to know how the scaling function $F(x,y)$ behaves for small $y$.
As already pointed out, this depends on the behavior of the integral at the upper limit
of integration, which is convergent (divergent) for $a < 1$ ($a \ge 1$). Hence,
from $1-a = (d_{\chi} - d)/2$ with $d_{\chi} = 4$ follows
\be
     F(x,y) \sim  \left \{ \begin{array}{ll}
        x^{-a}   \qquad $for$ \qquad d < d_{\chi}  \\
        x^{-a}\log (x/y)   \qquad $for$ \qquad d=d_{\chi} \\
       y^{1-a}/x        \qquad $for$ \qquad d > d_{\chi}. 
        \end{array}
        \right .
        \label{7.4}
        \ee
Inserting into Eq.~(\ref{7.2}) and comparing with Eq.~(\ref{3.4.3}) we recover
Eq.s~(\ref{4.1}) and (\ref{4.4}). Finally, for large $x$ we obtain the 
analogous of Eq.~(\ref{5.5})
\be
    \widehat{\chi}(x) \sim \left \{ \begin{array}{ll}
        x^{-a} \qquad $for$ \qquad d < d_{\chi}  \\
        x^{-a} \log x  \qquad $for$ \qquad d=d_{\chi} \\
	x^{-1}   \qquad $for$ \qquad d > d_{\chi}. 
        \end{array}
        \right .
        \label{3.5.1}
        \ee
Notice that the separation of the stationary from the aging response function has played
a crucial role. Had we used   
the form (\ref{2.14}) of HPGL in  Eq.~(\ref{7.3}) we would have obtained a completely
different behavior, with $d_{\chi}=2$ and in place of Eq.~(\ref{4.1}) 
\be
    a_{\chi} = \left \{ \begin{array}{ll}
        a= d/2-1  \qquad $for$ \qquad d < 2  \\
        0  \qquad $with log corrections for$ \qquad d=2 \\
	0   \qquad $for$ \qquad d > 2.
        \end{array}
        \right .
        \label{3.5.2}
        \ee

In order to illustrate the difference in the behaviors of TRM and ZFC we have solved
numerically for $\chi_{ag}(t,t_w)$ and for the corresponding effective
exponent $a_{\chi,eff}(t_w,x)$ (Fig.s~\ref{fig.10},\ref{fig.11}) with the same values of $d$ and $r$
used for TRM. These Figures show that both above and below 
$d_{\chi}$ there is no crossover, but there is only the early regime followed abruptly by
the asymptotic power law behavior, as for TRM above $d_{\rho}$ (Fig.~\ref{fig.9}).
Furthermore, we have depicted in Fig.s~\ref{fig.12},\ref{fig.13} the scaling function $\widehat{\chi}(x)$,
obtained by plotting $t_w^{ a_{\chi}}\chi_ {ag}(t_w,x)$ versus $x$ for different
$t_w$, which obeys the power laws (\ref{3.5.1}) for large $x$. These are the analogous 
of Fig.s~\ref{fig.3},\ref{fig.5}.

We can now summarise what we have learned from the large $N$ model about the connection 
between $a_{\rho}$, $a_{\chi}$ and $a$. In this case the explicit solution (\ref{2.18.1})
is available and $a$ is a linearly increasing function of dimensionality vanishing at $d_L=2$. 
The question is how much
of this could have been inferred relying only on the information from TRM or ZFC.
The answer is that both $a_{\rho}$ and $a_{\chi}$ coincide with $a$ below certain
dimensionalities $d_{\rho}$ and $d_{\chi}$.
At $d=d_{\rho}$ and $d=d_{\chi}$ there are logarithmic corrections, while above these 
dimensionalities  $a_{\rho}$ and $a_{\chi}$ are different from $a$ and differ one
from the other (Fig.~\ref{fig.14}). Although in the large $N$ model $d_{\rho} = d_{\chi} =4$, we
have kept distinct notations because $d_{\rho}$, which is the dimensionality where
$\lambda/z = a$, and  $d_{\chi}$ the dimensionality where $a-1=0$, need not to coincide in general.
In the large $N$ model they do coincide because of Eq.~(\ref{2.18.8}).
 Furthermore, even below $d_{\rho}$ and $d_{\chi}$, where $a_{\rho} = a_{\chi} =a$, there
remains a considerable difference between TRM and ZFC in relation to the time scales
($t^*$ and $t_{sc}$) over which these exponents are observable. Comparing Fig.s~\ref{fig.8},\ref{fig.10} 
one can see at glance that the difference between these time scales in
certain conditions, here set by the value of $r$, can become huge and if working with
TRM it may require an enormous $t_w$ before reaching  the asymptotic regime where
$a_{\rho}$ and $a$ can be identified.

\vspace{1cm}
\begin{figure}
\centerline{\psfig{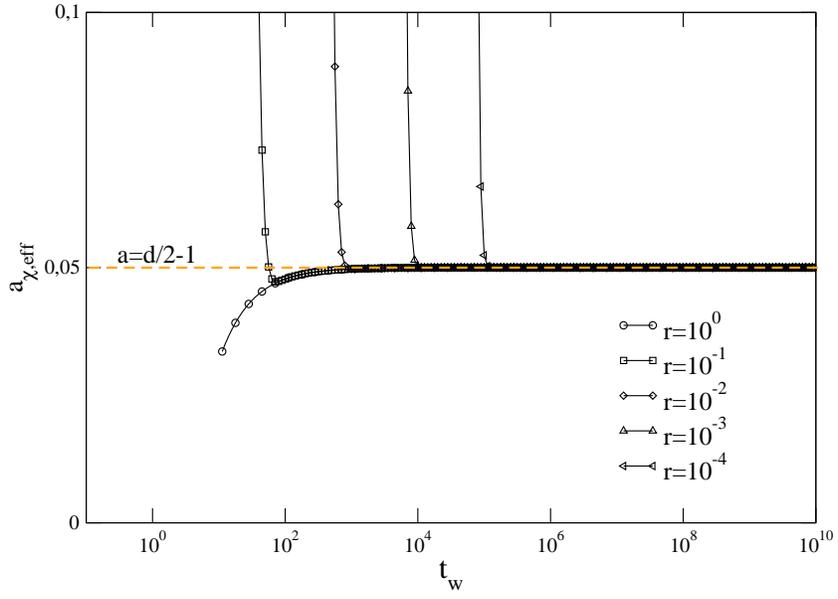}}
\caption{Effective exponent $a_{\chi,eff}$ in the large $N$ model 
versus $t_w$ for different values of $r$
with $x= 20$, $d=2.1$ and $T=0$.}
\label{fig.10}
\end{figure}

\vspace{1cm}
\begin{figure}
\centerline{\psfig{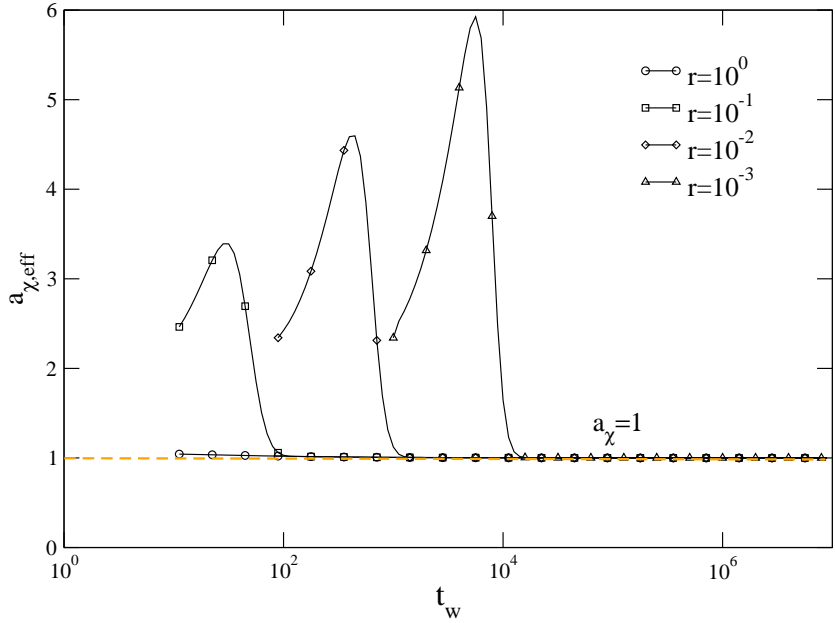}}
\caption{Effective exponent $a_{\chi,eff}$ in the large $N$ model 
versus $t_w$ for different values of $r$
with $x= 20$, $d=5$ and $T=0$.
}
\label{fig.11}
\end{figure}

\vspace{1.5cm}
\begin{figure}
\centerline{\psfig{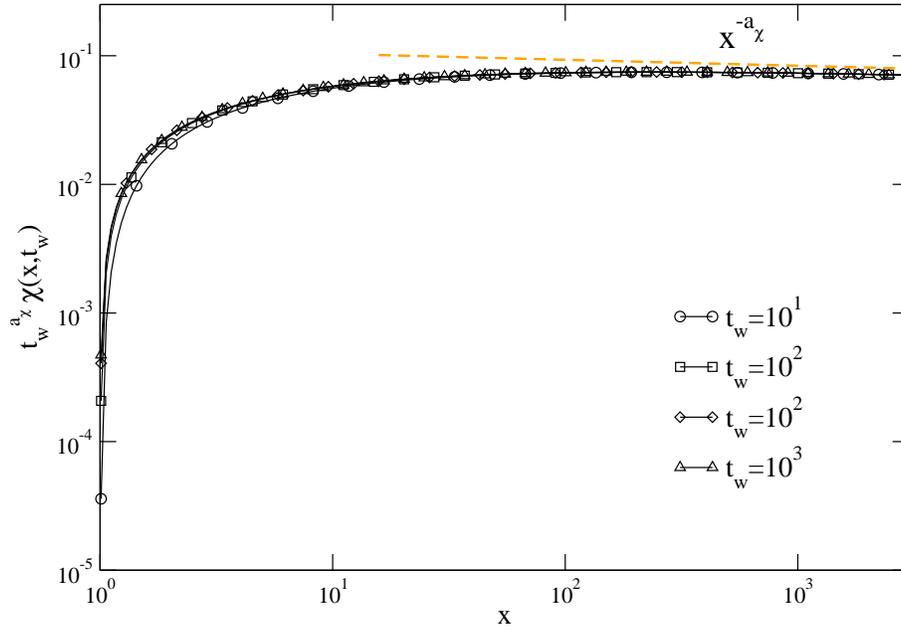}}
\caption{Scaling function $\widehat{\chi}(x)$ in the large $N$ model 
with $d=2.1$ and $T=0$. In this case $a_{\chi}=a=0.05$.}
\label{fig.12}
\end{figure}

\vspace{1cm}
\begin{figure}
\centerline{\psfig{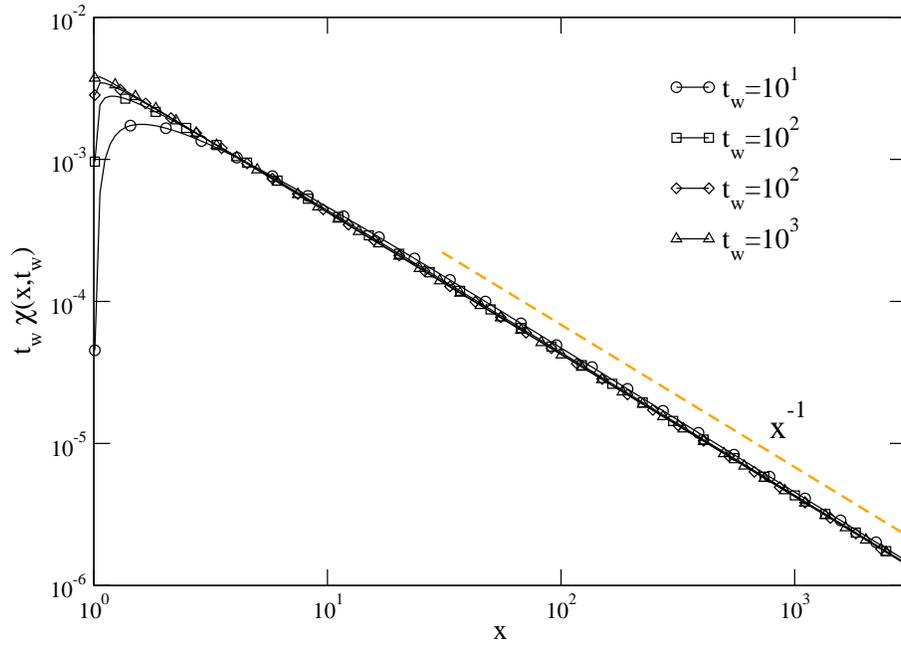}}
\caption{Scaling function $\widehat{\chi}(x)$ in the large $N$ model 
with  $d=5$ and $T=0$. In this case $a_{\chi}=1$ and $a=1.5$.}
\label{fig.13}
\end{figure}

\vspace{1.5cm}

\begin{figure}
\centerline{\psfig{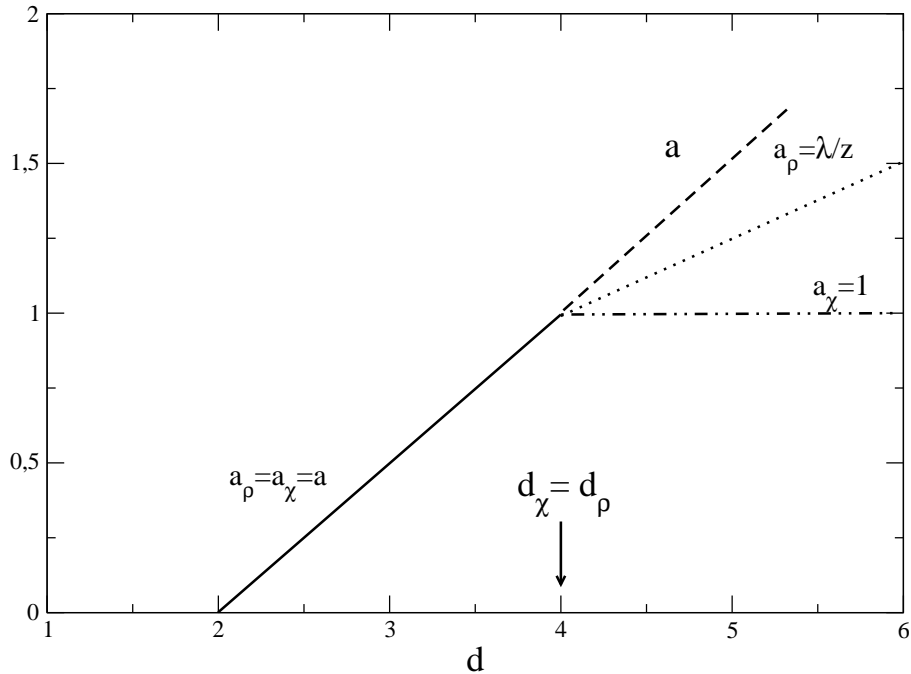}}
\caption{Overview of the dimensionality dependence of the exponents $a,a_{\chi},a_{\rho}$
in the large $N$ model.
}
\label{fig.14}
\end{figure}

\section{Concluding remarks and open problems} \label{concl}

In conclusion, we have shown that all existing analytical results and the numerical
evidence coming from ZFC in the Ising model are 
consistent with an exponent $a$ of the form (\ref{4.4}). The dimensionality
independent behavior (\ref{3.7}) predicted by the qualitative argument for
$a_{\chi}$ holds only for $d > d_{\chi}$ where $a_{\chi} \neq a$. This is 
due to the presence of a dangerous irrelevant variable. Once this is taken into
account, analytical and numerical results form a coherent picture and the
issue can be considered as settled.

For what concerns Eq.~(\ref{3.10}), regarded in  
Ref.s~\cite{Henkel2001,Henkel2003,HPM2003,HM2003} as  
the exponent $a$ in the Ising model, we have shown
that it does not have any analytical foundation, because Ref.~\cite{Berthier}
contains a computation of $a_{\chi}$. Furthermore, the numerical
evidence, being based on TRM data, is inconclusive since the largest $t_w$ 
reached so far are below the crossover time $t^*$. 
Therefore, $t_w$ is still far from being well
inside the asymptotic region as required for the TRM data to qualify as a
challange to those obtained from ZFC.
There is no doubt that among all possible IRF that one can employ to
study the exponent $a$, TRM is the most unfavourable and the less reliable one,
as abundantly explained in the previous sections. 

For what concerns the scaling function $f(x,y)$,
our ZFC data are consistent with an $f(x,y)$ in the Ising model 
of the form (\ref{2.15}) with the exponent $\alpha =a+1/2$ in place of $\alpha =a+1$,
appearing in the HPGL theory. We
have also shown that with the HPGL theory it is not possible to reproduce the
short time behavior of $R_{ag}(t,s,t_0)$. Nonetheless, our knowledge of the
scaling function $f(x,y)$ is still incomplete, since from ZFC data we cannot determine 
the exponent $\beta$.

After this survey of what can and what cannot be done with ZFC and TRM, 
it seems clear that in order to study $R_{ag}(t,s,t_0)$
the right thing to do would be to use neither of them. Rather, one should use
an IRF of the general form (\ref{3.1}) with $t_1 \gg t_{sc}$ 
to eliminate the crossover affecting TRM and with $t_2 < t$ in order
to avoid the dangerous irrelevant variable in ZFC. Namely, assuming
the form~(\ref{2.15}) of $f(x,y)$ and using Eq.~(\ref{3.1}) one should consider
\be
\mu_{ag}(t,t_2,t_1,t_0) = t^{-a} \int_{t_1/t}^{t_2/t} dz
\frac{z^{\beta +\alpha -1 -a}}{(1-z +t_0/t)^{\alpha}}.
\label{5.1}
\ee
If $t_2 <t$ and $t \gg t_0$, the dependence on $t_0$ can be neglected and
the above equation can be used in two ways. Rewriting
\be
\mu_{ag}(t,t_2,t_1) = t^{-a} \int_{x_1}^{x_2} dz \frac{z^{\beta +\alpha -1 -a}}{(1-z)^{\alpha}}
\label{5.2}
\ee
and keeping $x_1=t_1/t$ and $x_2=t_2/t$ fixed, the exponent
$a$ can be measured. 
Next, for $t \gg t_2$ from Eq. (\ref{5.1}) follows
\be
\mu_{ag}(t,t_2,t_1) \sim  t^{-a} \int_{t_1/t}^{t_2/t} dz z^{\beta +\alpha -1 -a}
\sim t^{-(\beta +\alpha)}
\label{5.4}
\ee
from which $\beta +\alpha$ can be measured, while $\alpha$ as we have seen can be 
extracted from $\widehat{\chi}(x)$. We plan to pursue the investigation of  this IRF 
in future work. 

Finally, the results obtained in this paper open a number of interesting problems 
in the general theory of phase ordering. We stress that our results are phenomenological.
In particular, we do not know why $d_L$ and  $d_{\chi}$ take the values they
take. $d_L$ seems to coincide with the ordinary lower critical dimensionality,
but we do not know whether this is really so, or it is just a coincidence. 
Even less we can tell about the values taken by the upper
dimensionality $d_{\chi}$. It should be noted the failure of the
GAF approximation to reproduce the correct dependence of $a$ on $d$ in the scalar case. 
In short, we have no theory for the observed behavior
of the response function in phase ordering kinetics.

\vspace{5mm}

{\bf Acknowledgments} - This work has been partially supported
from MURST through PRIN-2002.

\section{Appendix}

From the definitions (\ref{6.20}), (\ref{6.21}) and (\ref{6.22}) we have
\be 
t^* = \left [ \frac{B(t_{sc}) |\lambda/z -a|}{A}  \pm t_{sc}^{\lambda/z -a}\right ]^{1/(\lambda/z -a)}
\label{A1}
\ee
with the $+(-)$ sign if $\lambda/z -a <(>) 0$, i.e if $d >(<) 4$.
In order to estimate $B(t_{sc})$ we use the linear approximation 
\be
B(t_{sc})= \int_0^{t_{sc}} e^{rs}= (e^{rt_{sc}}-1)/r.
\label{A2}
\ee
Then, using  $A=(8 \pi)^{-d/2} \Delta/ M_0^2$ and $t_{sc}=-d/4r$
we find 
\be
B(t_{sc})/A=Ct_{sc}
\label{A2.0}
\ee
with 
\be
C=4(8 \pi)^{d/2}(1-e^{-d/4}) M_0^2/\Delta d
\label{A2.1}
\ee
and inserting
into Eq.~(\ref{A1}) we get
\be
t^*/t_{sc} = \left [ \frac{(4-d)}{4}C t_{sc}^{d/4} \pm 1 \right ]^{4/(4-d)}.
\label{A3}
\ee
For $d<4$ the above equation must be taken with the minus sign. This requires
$t_{sc} > [C(4-d)/4]^{-4/d}$ or $4 |r|/d < [C(4-d)/4]^{4/d}$. To lift this
restriction on the value of $r$ one must do better than the linear approximation in
the estimate of $B(t_{sc})$. Taking $t_{sc}$ large enough Eq.~(\ref{6.29}) is obtained.

\vspace{3mm}

If $d>4$, instead, from Eq.~(\ref{A3}) follows $t^* < t_{sc}$ justifying Eq.~(\ref{2.34.4}).

\vspace{3mm}

Finally, for $d=4$ from  Eq.~(\ref{2.18.0}) we get 
\be
\rho(t,t_w,t^*,t_{sc})= t_w^{-\lambda/z} \left [ \frac{B(t_{sc})}{A} + \log (t_w/t_{sc}) \right ]
(4 \pi)^{-d/2} (t/t_w)^{-\lambda/z}
\label{A4}
\ee
and defining $t^*$ by 
\be
\log (t^*/t_{sc}) = B(t_{sc})/A
\label{A2.2}
\ee
Eq.s~(\ref{6.32}) and (\ref{6.34}) are recovered after using Eq.~(\ref{A2.0}).

\end{document}